\documentclass[aps,preprint]{revtex4}%
\usepackage{amssymb}
\usepackage{amsmath}
\usepackage{amsfonts}
\usepackage{graphicx}%
\begin{document}
\title{Anatomy of Zero-norm States in String Theory}
\author{Chuan-Tsung Chan}
\email{ctchan@phys.cts.nthu.edu.tw}
\affiliation{Physics Division, National Center for Theoretical Sciences, Hsinchu, Taiwan, R.O.C.}
\author{Jen-Chi Lee}
\email{jcclee@cc.nctu.edu.tw}
\affiliation{Department of Electrophysics, National Chiao-Tung University, Hsinchu, Taiwan, R.O.C.}
\author{Yi-Yang}
\email{yiyang@mail.nctu.edu.tw}
\affiliation{Department of Electrophysics, National Chiao-Tung University, Hsinchu, Taiwan, R.O.C.}
\date{\today }

\begin{abstract}
We calculate and identify the counterparts of zero-norm states in the old
covariant first quantised (OCFQ) spectrum of open bosonic string in two other
quantization schemes of string theory, namely the light-cone DDF zero-norm
states and the off-shell BRST zero-norm states (with ghost) in the Witten
string field theory (WSFT). In particular, special attention is paid to the
inter-particle zero-norm states in all quantization schemes. For the case of
the off-shell BRST zero-norm states, we impose the no ghost conditions and
recover exactly two types of on-shell zero-norm states in the OCFQ string
spectrum for the first few low-lying mass levels. We then show that off-shell
gauge transformations of WSFT are identical to the on-shell stringy gauge
symmetries generated by two types of zero-norm states in the generalized
massive $\sigma$-model approach of string theory. The high energy limit of
these stringy gauge symmetries was recently used to calculate the
proportionality constants, conjectured by Gross, among high energy scattering
amplitudes of different string states. Based on these zero-norm state
calculations, we have thus related gauge symmetry of WSFT to the high-energy
stringy symmetry of Gross.

\end{abstract}
\maketitle

\section{Introduction}

Recently it was discovered that \cite{1,2} the high-energy limit
$\alpha^{\prime}\rightarrow\infty$ of stringy Ward identities, derived from
the decoupling of two types of zero-norm states in the OCFQ spectrum, imply an
infinite number of linear relations among high energy scattering amplitudes of
different string states with the same momenta. Moreover, these linear
relations can be used to fix the proportionality constants between high energy
scattering amplitudes of different string states algebraically at \textit{each
fixed mass level}. Thus there is only one independent component of high energy
scattering amplitude at each fixed mass level. For the case of string-tree
amplitudes, a general formula can even be given to express all high-energy
stringy scattering amplitudes for \textit{arbitrary mass levels} in terms of
those of tachyons \cite{1,3}. These zero-norm state calculations are
independent of the high-energy saddle-point calculations of Gross and Mende
\cite{4}, Gross \cite{5} and Gross and Manes \cite{6}. In fact, the results of
saddle-point calculations by those authors were found \cite{1,2,3} to be
inconsistent with high energy stringy Ward identities of zero-norm state
calculations, and thus could threat the validity of unitarity of string
perturbation theory. A corrected saddle-point calculation was given in
\cite{3}, where the missing terms of the calculations in Refs [4,5,6] were
identified to recover the stringy Ward identities.

The importance of zero-norm states and their implication on stringy symmetries
were first pointed out in the context of massive $\sigma$-model approach of
string theory \cite{7}. Some implications of the corresponding stringy Ward
identities on the scattering amplitudes were discussed in \cite{8}. On the
other hand, zero-norm states were also shown \cite{9} to carry the spacetime
$\omega_{\infty}$ symmetry \cite{10} charges of 2D string theory \cite{11}.
This is in parallel with the work of \cite{12} where the ground ring structure
of ghost number zero operators was identified in the BRST quantization. All
the above interesting results of 26D and 2D string theories strongly suggest
that a clearer understanding of zero-norm states holds promise to
\textit{uncover the fundamental symmetry of string theory.} Recently, a
simplified method to generate zero-norm states in 26D OCFQ bosonic string was
proposed \cite{13}. Based on a simplified prescription to calculate
positive-norm propagating states given in \cite{14}, general formulas of some
zero-norm tensor states at arbitrary mass levels were calculated.
Unfortunately, general formulas for the \textit{complete} set of zero-norm
states are still lacking mostly due to the high dimensionality of spacetime D
= 26. However, in the 2D OCFQ string theory, a general formula of zero-norm
states with discrete Polyakov's momenta at arbitrary mass levels was given in
terms of Schur Polynomials \cite{9}. On the other hand, for the case of 26D
string, the background ghost transformations in the gauge transformations of
WSFT \cite{15} were shown \cite{16} to correspond, in a one-to-one manner, to
the lifting of on-shell conditions of zero-norm states in the OCFQ approach.

In this paper, we shall calculate and identify the counterparts of zero-norm
states in two other quantization schemes of 26D open bosonic string theory,
namely the light-cone DDF \cite{17} zero-norm states and the off-shell BRST
zero-norm states (with ghost) in WSFT. In particular, special attention is
paid to the inter-particle zero-norm states in all quantization schemes. For
the case of off-shell BRST zero-norm states, we impose the no ghost conditions
and recover exactly two types of on-shell zero-norm states in the OCFQ string
spectrum for the first few low-lying mass levels. We then show that off-shell
gauge transformations of WSFT are identical to the on-shell stringy gauge
symmetries generated by two typse of zero-norm states in the OCFQ string
theory. Our calculations in this paper serve as the first step to study
stringy symmetries in light-cone DDF and BRST string theories, and to bridge
the links between different quantization schemes for both on-shell and
off-shell string theories. In section II, we first review the calculations of
zero-norm states in OCFQ spectrum. The most general spectrum analysis in the
helicity basis, including zero-norm states, was then given to discuss the
inter-particle $D_{2}$ zero-norm state \cite{7,8} at mass level $m^{2}=4$. We
will see that \textit{one can use polarization of either one of the two
positive-norm states to represent the polarization of the}
\textit{inter-particle zero-norm state. }This justifies how one can have the
inter-particle symmetry transformation for the two massive modes in the weak
field massive $\sigma$-model calculation derived previously \cite{7}. In
section III, we calculate both type I and type II zero-norm states in the
light-cone DDF string up to mass level $m^{2}=4$. In section IV, we first
calculate off-shell zero-norm states with ghosts from linearized gauge
transformation of WSFT. After imposing the no ghost conditions on these
zero-norm states, we can reproduce exactly two types of zero-norm states in
OCFQ spectrum for the first few low-lying mass levels. We then show that
off-shell gauge transformations of WSFT are identical to the on-shell stringy
gauge symmetries generated by two typse of zero-norm states in the generalized
massive $\sigma$-model approach \cite{7} of string theory. The high energy
limit of these stringy gauge symmetries was recently used to calculate the
proportionality constants among high energy scattering amplitudes of different
string states conjectured by Gross \cite{5}. \textit{Based on the zero-norm
state calculations \cite{1,2,3}, we have thus related gauge symmetry of WSFT
\cite{15} to the high-energy stringy symmetry conjectured by Gross
\cite{4,5,6}.} Finally, a brief conclusion is given in section V.

\section{\bigskip Zero-norm states in the OCFQ spectrum}

In the OCFQ spectrum of open bosonic string theory, the solutions of physical
states conditions include positive-norm propagating states and two types of
zero-norm states. The latter are \cite{18}
\begin{align}
\text{Type I}:L_{-1}\left\vert x\right\rangle ,  &  \text{ where }%
L_{1}\left\vert x\right\rangle =L_{2}\left\vert x\right\rangle =0,\text{
}L_{0}\left\vert x\right\rangle =0;\tag{2.1}\\
\text{Type II}:(L_{-2}+\frac{3}{2}L_{-1}^{2})\left\vert \widetilde
{x}\right\rangle ,  &  \text{ where }L_{1}\left\vert \widetilde{x}%
\right\rangle =L_{2}\left\vert \widetilde{x}\right\rangle =0,\text{ }%
(L_{0}+1)\left\vert \widetilde{x}\right\rangle =0. \tag{2.2}%
\end{align}
Eqs.(2.1) and (2.2) can be derived from Kac determinant in conformal field
theory. While type I states have zero-norm at any spacetime dimension, type II
states have zero-norm \textit{only} at D=26.

\subsection{Zero-norm states with constraints\bigskip}

The solutions of Eqs.(2.1) and (2.2) up to the mass level $m^{2}=4$ are listed
in the following \cite{13}:

1. $m^{2}=-k^{2}=0:$%

\begin{equation}
L_{-1\text{ }}\left\vert x\right\rangle =k\cdot\alpha_{-1}\left\vert
0,k\right\rangle ;\left\vert x\right\rangle =\left\vert 0,k\right\rangle
;\left\vert x\right\rangle =\left\vert 0,k\right\rangle . \tag{2.3}%
\end{equation}

2. $m^{2}=-k^{2}=2:$%

\begin{equation}
(L_{-2}+\frac{3}{2}L_{-1}^{2})\left\vert \widetilde{x}\right\rangle =[\frac
{1}{2}\alpha_{-1}\cdot\alpha_{-1}+\frac{5}{2}k\cdot\alpha_{-2}+\frac{3}%
{2}(k\cdot\alpha_{-1})^{2}]\left\vert 0,k\right\rangle ;\left\vert
\widetilde{x}\right\rangle =\left\vert 0,k\right\rangle , \tag{2.4}%
\end{equation}

\begin{equation}
L_{-1}\left\vert x\right\rangle =[\theta\cdot\alpha_{-2}+(k\cdot\alpha
_{-1})(\theta\cdot\alpha_{-1})]\left\vert 0,k\right\rangle ;\left\vert
x\right\rangle =\theta\cdot\alpha_{-1}\left\vert 0,k\right\rangle ,\theta\cdot
k=0. \tag{2.5}%
\end{equation}

3. $m^{2}=-k^{2}=4:$%

\begin{align}
(L_{-2}+\frac{3}{2}L_{-1}^{2})\left\vert \widetilde{x}\right\rangle  &
=\{4\theta\cdot\alpha_{-3}+\frac{1}{2}(\alpha_{-1}\cdot\alpha_{-1}%
)(\theta\cdot\alpha_{-1})+\frac{5}{2}(k\cdot\alpha_{-2})(\theta\cdot
\alpha_{-1})\nonumber\\[0.01in]
&  +\frac{3}{2}(k\cdot\alpha_{-1})^{2}(\theta\cdot\alpha_{-1})+3(k\cdot
\alpha_{-1})(\theta\cdot\alpha_{-2})\}\left\vert 0,k\right\rangle ;\nonumber\\
\left\vert \widetilde{x}\right\rangle  &  =\theta\cdot\alpha_{-1}\left\vert
0,k\right\rangle ,k\cdot\theta=0, \tag{2.6}%
\end{align}

\begin{align}
L_{-1}\left\vert x\right\rangle  &  =[2\theta_{\mu\nu}\alpha_{-1}^{\mu}%
\alpha_{-2}^{\nu}+k_{\lambda}\theta_{\mu\nu}\alpha_{-1}^{\lambda}\alpha
_{-1}^{\mu}\alpha_{-1}^{\nu}]\left\vert 0,k\right\rangle ;\nonumber\\
\left\vert x\right\rangle  &  =\theta_{\mu\nu}\alpha_{-1}^{\mu\nu}\left\vert
0,k\right\rangle ,k\cdot\theta=\eta^{\mu\nu}\theta_{\mu\nu}=0,\theta_{\mu\nu
}=\theta_{\nu\mu},\tag{2.7}%
\end{align}

\begin{align}
L_{-1}\left\vert x\right\rangle  &  =[\frac{1}{2}(k\cdot\alpha_{-1}%
)^{2}(\theta\cdot\alpha_{-1})+2\theta\cdot\alpha_{-3}+\frac{3}{2}(k\cdot
\alpha_{-1})(\theta\cdot\alpha_{-2})\nonumber\\
&  +\frac{1}{2}(k\cdot\alpha_{-2})(\theta\cdot\alpha_{-1})]\left\vert
0,k\right\rangle ;\text{ \ }\nonumber\\
\text{\ }\left\vert x\right\rangle  &  =[2\theta\cdot\alpha_{-2}+(k\cdot
\alpha_{-1})(\theta\cdot\alpha_{-1})]\left\vert 0,k\right\rangle ,\theta\cdot
k=0,\tag{2.8}%
\end{align}

\begin{align}
L_{-1}\left\vert x\right\rangle  &  =[\frac{17}{4}(k\cdot\alpha_{-1}%
)^{3}+\frac{9}{2}(k\cdot\alpha_{-1})(\alpha_{-1}\cdot\alpha_{-1}%
)+9(\alpha_{-1}\cdot\alpha_{-2})\nonumber\\
&  +21(k\cdot\alpha_{-1})(k\cdot\alpha_{-2})+25(k\cdot\alpha_{-3})]\left\vert
0,k\right\rangle ;\nonumber\\
\left\vert x\right\rangle  &  =[\frac{25}{2}k\cdot\alpha_{-2}+\frac{9}%
{2}\alpha_{-1}\cdot\alpha_{-1}+\frac{17}{4}(k\cdot\alpha_{-1})^{2}]\left\vert
0,k\right\rangle .\tag{2.9}%
\end{align}

Note that there are two degenerate vector zero-norm states, Eq.(2.6) for type
II and Eq.(2.8) for type I, at mass level $m^{2}=4$. We define $D_{2}$ vector
zero-norm state by antisymmetrizing those terms which contain $\alpha
_{-1}^{\mu}\alpha_{-2}^{\nu}$ in Eqs.(2.6) and (2.8) as following \cite{7}%

\begin{equation}
|D_{2}\rangle=[(\frac{1}{2}k_{\mu}k_{\nu}\theta_{\lambda}+2\eta_{\mu\nu}%
\theta_{\lambda})\alpha_{-1}^{\mu}\alpha_{-1}^{\nu}\alpha_{-1}^{\lambda
}+9k_{\mu}\theta_{\nu}\alpha_{-2}^{[\mu}\alpha_{-1}^{\nu]}-6\theta_{\mu}%
\alpha_{-3}^{\mu}]\left\vert 0,k\right\rangle ,\text{ \ }k\cdot\theta=0.
\tag{2.10}%
\end{equation}
Similarly $D_{1}$ vector zero-norm state is defined by symmetrizing those
terms which contain $\alpha_{-1}^{\mu}\alpha_{-2}^{\nu}$ in Eqs.(2.6) and
(2.8)
\begin{equation}
|D_{1}\rangle=[(\frac{5}{2}k_{\mu}k_{\nu}\theta_{\lambda}+\eta_{\mu\nu}%
\theta_{\lambda})\alpha_{-1}^{\mu}\alpha_{-1}^{\nu}\alpha_{-1}^{\lambda
}+9k_{\mu}\theta_{\nu}\alpha_{-2}^{(\mu}\alpha_{-1}^{\nu)}+6\theta_{\mu}%
\alpha_{-3}^{\mu}]\left\vert 0,k\right\rangle \text{, \ }k\cdot\theta=0.
\tag{2.11}%
\end{equation}
\qquad\qquad

In the generalized massive $\sigma$-model approach of string theory, it can be
shown that each zero-norm state in the OCFQ spectrum generates a massive
symmetry transformation for the propagating string modes. In particular, the
inter-particle symmetry transformation corresponding to the $D_{2}$
\textit{inter-particle zero-norm state} in Eq.(2.10) can be calculated to be
\cite{7}%

\begin{equation}
\delta C_{(\mu\nu\lambda)}=(\frac{1}{2}\partial_{(\mu}\partial_{\nu}%
\theta_{\lambda)}-2\eta_{(\mu\nu}\theta_{\lambda)}),\delta C_{[\mu\nu
]}=9\partial_{\lbrack\mu}\theta_{\nu]}, \tag{2.12}%
\end{equation}
where $\partial_{\nu}\theta^{\nu}=0,(\partial^{2}-4)\theta^{\nu}=0$ are the
on-shell conditions of the $D_{2}$ vector zero-norm state. $C_{(\mu\nu
\lambda)}$ and $C_{[\mu\nu]}$ in Eq.(2.12) are the background fields of the
symmetric spin-three and antisymmetric spin-two states, respectively, at mass
level $m^{2}=4$. Eq.(2.12) is the result of the first order weak field
approximation but, in contrast to the high energy $\alpha^{\prime}%
\rightarrow\infty$ result of \cite{1,2,3}, valid to \emph{all} energy
$\alpha^{\prime}$ in the generalized $\sigma$-model approach. It is important
to note that the decoupling of $D_{2}$ vector zero-norm state implies
simultaneous change of both $C_{(\mu\nu\lambda)}$ and $C_{[\mu\nu]}$ , thus
they form a gauge multiplet. In general, \textit{an inter-particle zero-norm
state} can be defined to be $D_{2}+\alpha D_{1}$, where $\alpha$ is an
arbitrary constant.

\subsection{\bigskip Zero-norm states in the helicity basis}

In this subsection, we are going to do the most general spectrum analysis
which naturally includes zero-norm states. We will then \textit{solve} the
Virasoro constraints in the helicity basis and recover the zero-norm states
listed above \cite{7,13}. In particular, this analysis will make it clear how
$D_{2\text{ }}$zero-norm state in Eq.(2.10) can induce the inter-particular
symmetry transformation for two propagating states at the mass level $m^{2}=4$.

We begin our discussion for the mass level $m^{2}=2$. At this mass level, the
general expression for the physical states can be written as
\begin{equation}
\lbrack\epsilon_{\mu\nu}\alpha_{-1}^{\mu}\alpha_{-1}^{\nu}+\epsilon_{\mu
}\alpha_{-2}^{\mu}]|0,k\rangle. \tag{2.13}%
\end{equation}
In the OCFQ of string theory, physical states satisfy the mass shell
condition
\begin{equation}
(L_{0}-1)|phys\rangle=0\Rightarrow k^{2}=-2; \tag{2.14}%
\end{equation}
and the Virasoro constraints $L_{1}|phys\rangle=L_{2}|phys\rangle=0$ which
give
\begin{align}
\epsilon_{\mu}  &  =-\epsilon_{\mu\nu}k^{\nu},\tag{2.15}\\
\eta^{\mu\nu}\epsilon_{\mu\nu}  &  =2\epsilon_{\mu\nu}k^{\mu}k^{\nu}.
\tag{2.16}%
\end{align}

In order to solve for the constraints Eq.(2.15) and Eq.(2.16) in a covariant
way, it is convenient to make the following change of basis,
\begin{align}
e_{P}  &  \equiv\frac{1}{m}(E,0,....,\text{k})\tag{2.17}\\
e_{L}  &  \equiv\frac{1}{m}(\text{k},0,....,E)\tag{2.18}\\
e_{T_{i}}  &  \equiv(0,0,....,1(\text{i-th spatial direction}),....,0),\hspace
{0.5cm}i=1,2,....,24. \tag{2.19}%
\end{align}
The 2nd rank tensor $\epsilon_{\mu\nu}$ can be written in the helicity basis
Eqs.(2.17)-(2.19) as
\begin{equation}
\epsilon_{\mu\nu}=\sum_{A,B}u_{AB}e_{\mu}^{A}e_{\nu}^{B},\hspace
{1.5cm}A,B=P,L,T_{i}. \tag{2.20}%
\end{equation}

In this new representation, the second Virasoro constraint Eq.(2.16) reduces
to a simple algebraic relation, and one can solve it
\begin{equation}
u_{PP}=\frac{1}{5}(u_{LL}+\sum_{i=1}^{24}u_{T_{i}T_{i}}).\tag{2.21}%
\end{equation}
In order to perform an irreducible decomposition of the spin-two state into
the trace and traceless parts, we define the following variables
\begin{align}
x &  \equiv\frac{1}{25}(u_{LL}+\sum_{i=1}^{24}u_{T_{i}T_{i}}),\tag{2.22}\\
y &  \equiv\frac{1}{25}(u_{LL}-\frac{1}{24}\sum_{i=1}^{24}u_{T_{i}T_{i}%
}).\tag{2.23}%
\end{align}
We can then write down the complete decompositions of the spin-two
polarization tensor as
\begin{align}
\epsilon_{\mu\nu} &  =x\ (5e_{\mu}^{P}e_{\nu}^{P}+e_{\mu}^{L}e_{\nu}^{L}%
+\sum_{i=1}^{24}e_{\mu}^{T_{i}}e_{\nu}^{T_{i}})\nonumber\\
&  +y\ \sum_{i=1}^{24}(e_{\mu}^{L}e_{\nu}^{L}-e_{\mu}^{T_{i}}e_{\nu}^{T_{i}%
})\nonumber\\
&  +\sum_{i,j}(u_{T_{i}T_{j}}-\frac{\delta_{ij}}{24}\sum_{l=1}^{24}%
u_{T_{l}T_{l}})e_{\mu}^{T_{i}}e_{\nu}^{T_{j}}\nonumber\\
&  +u_{PL}(e_{\mu}^{P}e_{\nu}^{L}+e_{\mu}^{L}e_{\nu}^{P})\nonumber\\
&  +\sum_{i=1}^{24}u_{PT_{i}}(e_{\mu}^{P}e_{\nu}^{T_{i}}+e_{\mu}^{T_{i}}%
e_{\nu}^{P})\nonumber\\
&  +\sum_{i=1}^{24}u_{LT_{i}}(e_{\mu}^{L}e_{\nu}^{T_{i}}+e_{\mu}^{T_{i}}%
e_{\nu}^{L}).\tag{2.24}%
\end{align}
The first Virasoro constraint Eq.(2.15) implies that $\epsilon_{\mu}$ vector
is not an independent variable, and is related to the spin-two polarization
tensor $\epsilon_{\mu\nu}$ as follows
\begin{equation}
\epsilon_{\mu}=5\sqrt{2}xe_{\mu}^{P}+\sqrt{2}u_{PL}e_{\mu}^{L}+\sqrt{2}%
\sum_{i=1}^{24}u_{PT_{i}}e_{\mu}^{T_{i}}.\tag{2.25}%
\end{equation}
Finally, combining the results of Eqs.(2.13),(2.24) and (2.25), we get the
complete solution for physical states at mass level $m^{2}=2$
\begin{align}
&  [\epsilon_{\mu\nu}\alpha_{-1}^{\mu}\alpha_{-1}^{\nu}+\epsilon_{\mu}%
\alpha_{-2}^{\mu}]|0,k\rangle\nonumber\\
&  =x(5\alpha_{-1}^{P}\alpha_{-1}^{P}+\alpha_{-1}^{L}\alpha_{-1}^{L}%
+\sum_{i=1}^{24}\alpha_{-1}^{T_{i}}\alpha_{-1}^{T_{i}}+5\sqrt{2}\alpha
_{-2}^{P})|0,k\rangle\tag{2.26}\\
&  +y\sum_{i=1}^{24}(\alpha_{-1}^{L}\alpha_{-1}^{L}-\alpha_{-1}^{T_{i}}%
\alpha_{-1}^{T_{i}})|0,k\rangle\tag{2.27}\\
&  +\sum_{i,j}(u_{T_{i}T_{j}}-\frac{\delta_{ij}}{24}\sum_{l=1}^{24}%
u_{T_{l}T_{l}})\alpha_{-1}^{T_{i}}\alpha_{-1}^{T_{j}}|0,k\rangle\tag{2.28}\\
&  +u_{PL}(2\alpha_{-1}^{P}\alpha_{-1}^{L}+\sqrt{2}\alpha_{-2}^{L}%
)|0,k\rangle\tag{2.29}\\
&  +\sum_{i=1}^{24}u_{PT_{i}}(2\alpha_{-1}^{P}\alpha_{-1}^{T_{i}}+\sqrt
{2}\alpha_{-2}^{T_{i}})|0,k\rangle\tag{2.30}\\
&  +2\sum_{i=1}^{24}u_{LT_{i}}\alpha_{-1}^{L}\alpha_{-1}^{T_{i}}%
|0,k\rangle,\tag{2.31}%
\end{align}
where the oscillator creation operators $\alpha_{-1}^{P},\alpha_{-1}%
^{L},\alpha_{-1}^{T_{i}},$ etc., are defined as
\begin{equation}
\alpha_{-n}^{A}\equiv e_{\mu}^{A}\cdot\alpha_{-n}^{\mu},\hspace{1cm}n\in
N,\hspace{1cm}A=P,L,T_{i}.\tag{2.32}%
\end{equation}
In comparison with the standard expressions for zero-norm states in subsection
A, we find that Eqs. (2.26), (2.29) and (2.30) are identical to the type II
singlet and type I vector zero-norm states for the mass level $m^{2}$ $=2$%
\begin{align}
(2.26) &  =2x[(\frac{1}{2}\eta_{\mu\nu}+\frac{3}{2}k_{\mu}k_{\nu})\alpha
_{-1}^{\mu}\alpha_{-1}^{\nu}+\frac{5}{2}k_{\mu}\alpha_{-2}^{\mu}%
]|0,k\rangle,\nonumber\\
(2.29) &  =\sqrt{2}u_{PL}[e_{\mu}^{L}k_{\nu}\alpha_{-1}^{\mu}\alpha_{-1}^{\nu
}+e_{\mu}^{L}\alpha_{-2}^{\mu}]|0,k\rangle,\nonumber\\
(2.30) &  =\sum_{i=1}^{24}\sqrt{2}u_{PT_{i}}[e_{\mu}^{T_{i}}k_{\nu}\alpha
_{-1}^{\mu}\alpha_{-1}^{\nu}+e_{\mu}^{T_{i}}\alpha_{-2}^{\mu}]|0,k\rangle
.\tag{2.33}%
\end{align}
In addition, one can clearly see from our covariant decomposition how
zero-norm states generate gauge transformations on positve-norm states. While
a nonzero value for $x$ induces a gauge transformation along the type II
singlet zero-norm state direction, the coefficients $u_{PL},u_{PT_{i}}$
parametrize the type I vector gauge transformations with polarization vectors
$\theta=e^{L}$ and $\theta=e^{T_{i}}$, respectively. Finally, by a simple
counting of degrees of freedom, one can identify Eqs (2.27), (2.28) and (2.31)
as the singlet (1), (traceless) tensor (299), and vector (24) positive-norm
states, respectively. These positive-norm states are in a one-to-one
correspondence with the degrees of freedom in the light-cone quantization scheme.

We now turn to the analysis of $m^{2}=4$ spectrum. Due to the complexity of
our calculations, we shall present the calculations in three steps. We shall
first write down all of physical states (including both positive-norm and
zero-norm states) in the simplest gauge choices in the helicity basis. We then
calculate the spin-3 state decomposition in the most general gauge choice.
Finally, the complete analysis will be given to see how $D_{2\text{ }}%
$zero-norm state in Eq.(2.10) can induce the inter-particle symmetry
transformation for two propagating states at the mass level $m^{2}=4$.

\subsubsection{Physical states in the simplest gauge choices}

To begin with, let us first analyse the positive-norm states. There are two
particles at the mass level $m^{2}=4$, a totally symmetric spin-three particle
and an antisymmetric spin-two particle. The canonical representation of the
spin-three state is usually choosen as
\begin{equation}
\epsilon_{\mu\nu\lambda}\ \alpha_{-1}^{\mu}\alpha_{-1}^{\nu}\alpha
_{-1}^{\lambda}|0,k\rangle,\hspace{1cm}k^{2}=-4,\tag{2.34}%
\end{equation}
where the totally symmetric polarization tensor $\epsilon_{\mu\nu\lambda}$ can
be expanded in the helicity basis as
\begin{equation}
\epsilon_{\mu\nu\lambda}=\sum_{A,B,C}\ \tilde{u}_{ABC}e_{\mu}^{A}e_{\nu}%
^{B}e_{\lambda}^{C},\hspace{1cm}A,B,C=P,L,T_{i}.\tag{2.35}%
\end{equation}
The Virasoro conditions on the polarization tensor can be solved as follows%

\begin{align}
k^{\lambda}\epsilon_{\mu\nu\lambda}=0 &  \Rightarrow\tilde{u}_{PAB}%
=0,\hspace{0.5cm}\forall A,B=P,L,T_{i},\tag{2.36}\\
\eta^{\nu\lambda}\epsilon_{\mu\nu\lambda}=0 &  \Rightarrow\tilde{u}_{LLL}%
+\sum_{i}\tilde{u}_{T_{i}T_{i}L}=0,\nonumber\\
&  \tilde{u}_{LLT_{i}}+\sum_{j}\tilde{u}_{T_{j}T_{j}T_{i}}=0.\tag{2.37}%
\end{align}
If we choose to keep the minimal number of $L$ components in the expansion
coefficients $\tilde{u}_{ABC}$ for the spin-three particle, we get the
following canonical decomposition
\begin{align}
|A(\epsilon)\rangle\equiv( &  \epsilon_{\mu\nu\lambda}\alpha_{-1}^{\mu}%
\alpha_{-1}^{\nu}\alpha_{-1}^{\lambda})|0,k\rangle=|A(\tilde{u})\rangle
\nonumber\\
&  =\sum_{i}\tilde{u}_{T_{i}T_{i}T_{i}}(\alpha_{-1}^{T_{i}}\alpha_{-1}^{T_{i}%
}\alpha_{-1}^{T_{i}}-3\alpha_{-1}^{L}\alpha_{-1}^{L}\alpha_{-1}^{T_{i}%
})|0,k\rangle\nonumber\\
&  +\sum_{i\neq j}\ 3\ \tilde{u}_{T_{j}T_{j}T_{i}}(\alpha_{-1}^{T_{j}}%
\alpha_{-1}^{T_{j}}\alpha_{-1}^{T_{i}}-\alpha_{-1}^{L}\alpha_{-1}^{L}%
\alpha_{-1}^{T_{i}})|0,k\rangle\nonumber\\
&  +\sum_{(i\neq j\neq k)}6\ \tilde{u}_{T_{i}T_{j}T_{k}}(\alpha_{-1}^{T_{i}%
}\alpha_{-1}^{T_{j}}\alpha_{-1}^{T_{k}})|0,k\rangle\nonumber\\
&  +\sum_{i}\tilde{u}_{LT_{i}T_{i}}(3\ \alpha_{-1}^{L}\alpha_{-1}^{T_{i}%
}\alpha_{-1}^{T_{i}}-\alpha_{-1}^{L}\alpha_{-1}^{L}\alpha_{-1}^{L}%
)|0,k\rangle\nonumber\\
&  +\sum_{(i\neq j)}6\ \tilde{u}_{LT_{i}T_{j}}(\alpha_{-1}^{L}\alpha
_{-1}^{T_{i}}\alpha_{-1}^{T_{j}})|0,k\rangle.\tag{2.38}%
\end{align}
It is easy to check that the $2900$ independent degrees of freedom of the
spin-three particle decompose into $24+552+2024+24+276$ in the above representation.

Similarily, for the antisymmetric spin-two particle, we have the following
canonical representation
\begin{equation}
\epsilon_{\lbrack\mu,\nu]}\alpha_{-1}^{\mu}\alpha_{-2}^{\nu}|0,k\rangle.
\tag{2.39}%
\end{equation}
Rewriting the polarization tensor $\epsilon_{\lbrack\mu,\nu]}$ in the helicity
basis
\begin{equation}
\epsilon_{\lbrack\mu,\nu]}=\sum_{A,B}v_{[A,B]}e_{\mu}^{A}e_{\nu}^{B},
\tag{2.40}%
\end{equation}
and solving the Virasoro constraints
\begin{equation}
k^{\nu}\epsilon_{\lbrack\mu,\nu]}=2v_{[P,L]}e_{\mu}^{L}+2\sum_{i=1}%
^{24}v_{[P,T_{i}]}e_{\mu}^{T_{i}}=0, \tag{2.41}%
\end{equation}
we obtain the following decomposition for the spin-two state
\begin{align}
|B(\epsilon)\rangle\equiv &  \epsilon_{\lbrack\mu,\nu]}\alpha_{-1}^{\mu}%
\alpha_{-2}^{\nu}|0,k\rangle=|B(v)\rangle\nonumber\\
&  =\sum_{i}v_{[L,T_{i}]}(\alpha_{-1}^{L}\alpha_{-2}^{T_{i}}-\alpha
_{-1}^{T_{i}}\alpha_{-2}^{L})|0,k\rangle\nonumber\\
&  +\sum_{(i\neq j)}v_{[T_{i},T_{j}]}(\alpha_{-1}^{T_{i}}\alpha_{-2}^{T_{j}%
}-\alpha_{-1}^{T_{j}}\alpha_{-2}^{T_{i}})|0,k\rangle. \tag{2.42}%
\end{align}
Finally, one can check that the $300$ independent degrees of freedom of the
spin-two particle decompose into $24+276$ in the above expression.

For the zero-norm states at $m^{2}=4$, we have the following decompositions

\begin{enumerate}
\item Spin-two tensor
\begin{align}
|C(\theta)\rangle &  \equiv(k_{\lambda}\theta_{\mu\nu}\alpha_{-1}^{\mu}%
\alpha_{-1}^{\nu}\alpha_{-1}^{\lambda}+2\ \theta_{\mu\nu}\alpha_{-1}^{\mu
}\alpha_{-2}^{\nu})|0,k\rangle\nonumber\\
&  =\sum_{i}2\ \theta_{T_{i}T_{i}}(\alpha_{-1}^{T_{i}}\alpha_{-1}^{T_{i}%
}\alpha_{-1}^{P}-\alpha_{-1}^{L}\alpha_{-1}^{L}\alpha_{-1}^{P}+\alpha
_{-1}^{T_{i}}\alpha_{-2}^{T_{i}}-\alpha_{-1}^{L}\alpha_{-2}^{L})|0,k\rangle
\nonumber\\
&  +\sum_{(i\neq j)}2\ \theta_{T_{i}T_{j}}(2\ \alpha_{-1}^{T_{i}}\alpha
_{-1}^{T_{j}}\alpha_{-1}^{P}+\alpha_{-1}^{T_{i}}\alpha_{-2}^{T_{j}}%
+\alpha_{-1}^{T_{j}}\alpha_{-2}^{T_{i}})|0,k\rangle\nonumber\\
&  +\sum_{i}2\ \theta_{LT_{i}}(2\ \alpha_{-1}^{L}\alpha_{-1}^{T_{i}}%
\alpha_{-1}^{P}+\alpha_{-1}^{L}\alpha_{-2}^{T_{i}}+\alpha_{-1}^{T_{i}}%
\alpha_{-2}^{L})|0,k\rangle, \tag{2.43}%
\end{align}
where we have solved the Virasoro constraints on the polarization tensor
$\theta_{\mu\nu}$
\begin{align}
\theta_{\mu\nu}  &  =\sum_{A,B}\theta_{AB}e_{\mu}^{A}e_{\nu}^{B},\tag{2.44}\\
\eta^{\mu\nu}\theta_{\mu\nu}  &  =-\theta_{PP}+\theta_{LL}+\sum_{i}%
\theta_{T_{i}T_{i}}=0,\tag{2.45}\\
k^{\nu}\theta_{\mu\nu}  &  =-2\theta_{PP}e_{\mu}^{P}-2\theta_{PL}e_{\nu}%
^{L}-2\sum_{i}\theta_{PT_{i}}e_{\mu}^{T_{i}}=0. \tag{2.46}%
\end{align}
The $324$ degrees of freedom of on-shell $\theta_{\mu\nu}$decompose into
$24+276+24$ in Eq.(2.43).

\item Spin-one vector (with polarization vector $\theta\cdot k=0$,
$\theta_{\mu}=$ $\sum_{A}\theta_{A}e_{\mu}^{A},\ A$ $=L,T_{i}$)
\begin{align}
|D_{1}(\theta)\rangle &  \equiv\lbrack(\frac{5}{2}k_{\mu}k_{\nu}%
\theta_{\lambda}+\eta_{\mu\nu}\theta_{\lambda})\alpha_{-1}^{\mu}\alpha
_{-1}^{\nu}\alpha_{-1}^{\lambda}\nonumber\\
&  +9k_{(\mu}\theta_{\nu)}\alpha_{-1}^{\mu}\alpha_{-2}^{\nu}+6\theta_{\mu
}\alpha_{-3}^{\mu}]|0,k\rangle\tag{2.47}\\
&  =\sum_{A}\theta_{A}[9\alpha_{-1}^{P}\alpha_{-1}^{P}\alpha_{-1}^{A}%
+\alpha_{-1}^{L}\alpha_{-1}^{L}\alpha_{-1}^{A}+\sum_{i}\alpha_{-1}^{T_{i}%
}\alpha_{-1}^{T_{i}}\alpha_{-1}^{A}\nonumber\\
&  +9(\alpha_{-1}^{P}\alpha_{-2}^{A}+\alpha_{-1}^{A}\alpha_{-2}^{P}%
)+6\alpha_{-3}^{A}]|0,k\rangle. \tag{2.48}%
\end{align}

\item Spin-one vector (with polarization vector $\theta\cdot k=0$,
$\theta_{\mu}=$ $\sum_{A}\theta_{A}e_{\mu}^{A},\ A$ $=L,T_{i}$)%

\begin{align}
|D_{2}(\theta)\rangle &  \equiv\lbrack(\frac{1}{2}k_{\mu}k_{\nu}%
\theta_{\lambda}+2\eta_{\mu\nu}\theta_{\lambda})\alpha_{-1}^{\mu}\alpha
_{-1}^{\nu}\alpha_{-1}^{\lambda}\nonumber\\
&  -9k_{[\mu}\theta_{\nu]}\alpha_{-1}^{\mu}\alpha_{-2}^{\nu}-6\theta_{\mu
}\alpha_{-3}^{\mu}]|0,k\rangle\tag{2.49}\\
&  =\sum_{A}\theta_{A}[2\alpha_{-1}^{L}\alpha_{-1}^{L}\alpha_{-1}^{A}%
+2\sum_{j}\alpha_{-1}^{T_{j}}\alpha_{-1}^{T_{j}}\alpha_{-1}^{A}\nonumber\\
&  -9(\alpha_{-1}^{P}\alpha_{-2}^{A}-\alpha_{-1}^{A}\alpha_{-2}^{P}%
)-6\alpha_{-3}^{A}]|0,k\rangle. \tag{2.50}%
\end{align}

\item spin-zero singlet
\begin{align}
|E\rangle &  \equiv\lbrack(\frac{17}{4}k_{\mu}k_{\nu}k_{\lambda}+\frac{9}%
{2}\eta_{\mu\nu}k_{\lambda})\alpha_{-1}^{\mu}\alpha_{-1}^{\nu}\alpha
_{-1}^{\lambda}\nonumber\\
&  +(21k_{\mu}k_{\nu}+9\eta_{\mu\nu})\alpha_{-1}^{\mu}\alpha_{-2}^{\nu
}+25k_{\mu}\alpha_{-3}^{\mu}]|0,k\rangle\tag{2.51}\\
&  =[25(\alpha_{-1}^{P}\alpha_{-1}^{P}\alpha_{-1}^{P}+3\alpha_{-1}^{P}%
\alpha_{-2}^{P}+2\alpha_{-3}^{P})\nonumber\\
&  +9\alpha_{-1}^{L}\alpha_{-1}^{L}\alpha_{-1}^{P}+9\alpha_{-1}^{L}\alpha
_{-2}^{L}+9\sum_{i}(\alpha_{-1}^{T_{i}}\alpha_{-1}^{T_{i}}\alpha_{-1}%
^{P}+\alpha_{-1}^{T_{i}}\alpha_{-2}^{T_{i}})]|0,k\rangle. \tag{2.52}%
\end{align}

\end{enumerate}

\subsubsection{ \bigskip Spin-three state in the most general gauge choice}

In this subsection, we study the most general gauge choice associated with the
totally symmetric spin-three state
\begin{equation}
\lbrack\varepsilon_{\mu\nu\lambda}\alpha_{-1}^{\mu}\alpha_{-1}^{\nu}%
\alpha_{-1}^{\lambda}+\varepsilon_{(\mu\nu)}\alpha_{-1}^{\mu}\alpha_{-2}^{\nu
}+\varepsilon_{\mu}\alpha_{-3}^{\mu}]|0,k\rangle,\tag{2.53}%
\end{equation}
where Virasoro constraints imply
\begin{align}
\varepsilon_{(\mu\nu)} &  =-\frac{3}{2}k^{\lambda}\varepsilon_{\mu\nu\lambda
},\tag{2.54}\\
\varepsilon_{\mu} &  =\frac{1}{2}k^{\nu}k^{\lambda}\varepsilon_{\mu\nu\lambda
},\tag{2.55}\\
2\eta^{\mu\nu}\varepsilon_{\mu\nu\lambda} &  =k^{\mu}k^{\nu}\varepsilon
_{\mu\nu\lambda}.\tag{2.56}%
\end{align}
Eqs.(2.54) and (2.55) imply that both $\varepsilon_{(\mu\nu)}$ and
$\varepsilon_{\mu}$ are not independent variables, and Eq.(2.56) stands for
the constraint on the polarization $\varepsilon_{\mu\nu\lambda}$. In the
helicity basis, we define
\begin{equation}
\varepsilon_{\mu\nu\lambda}=\sum_{A,B,C}u_{ABC}\ e_{\mu}^{A}e_{\nu}%
^{B}e_{\lambda}^{C},\hspace{1cm}A,B,C=P,L,T_{i}.\tag{2.57}%
\end{equation}
Eq.(2.56) then gives
\begin{equation}
\sum_{A,B}\eta^{AB}u_{ABC}=2u_{PPC},\hspace{1cm}A,B,C=P,L,T_{i},\tag{2.58}%
\end{equation}
which implies%
\begin{equation}
3u_{PPC}-u_{LLC}-\sum_{j}u_{T_{j}T_{j}C}=0,\hspace{1cm}C=P,L,T_{i}.\tag{2.59}%
\end{equation}
Eliminating $u_{LLP},u_{LLL}$ and $u_{LLT_{i}}$ from above equations, we have
the solution for $\varepsilon_{\mu\nu\lambda}$, $\varepsilon_{(\mu\nu)}$ and
$\varepsilon_{\mu}$
\begin{align}
\varepsilon_{\mu\nu\lambda} &  =u_{PPP}\ [e_{\mu}^{P}e_{\nu}^{P}e_{\lambda
}^{P}+3(e_{\mu}^{L}e_{\nu}^{L}e_{\lambda}^{P}+\text{per.})]\nonumber\\
&  +u_{PPL}\ [(e_{\mu}^{P}e_{\nu}^{P}e_{\lambda}^{L}+\text{per.})+3e_{\mu}%
^{L}e_{\nu}^{L}e_{\lambda}^{L}]\nonumber\\
&  +\sum_{i}u_{PPT_{i}}\ [(e_{\mu}^{P}e_{\nu}^{P}e_{\lambda}^{T_{i}%
}+\text{per.})+3(e_{\mu}^{L}e_{\nu}^{L}e_{\lambda}^{T_{i}}+\text{per.}%
)]\nonumber\\
&  +\sum_{i}u_{PT_{i}T_{i}}\ [(e_{\mu}^{P}e_{\nu}^{T_{i}}e_{\lambda}^{T_{i}%
}+\text{per.})-(e_{\mu}^{L}e_{\nu}^{L}e_{\lambda}^{P}+\text{per.})]\nonumber\\
&  +\sum_{(i\neq j)}u_{PT_{i}T_{j}}\ [e_{\mu}^{P}e_{\nu}^{T_{i}}e_{\lambda
}^{T_{j}}+\text{per.}]\nonumber\\
&  +\sum_{i}u_{PLT_{i}}\ [e_{\mu}^{P}e_{\nu}^{L}e_{\lambda}^{T_{i}%
}+\text{per.}]\nonumber\\
&  +\sum_{i}u_{LT_{i}T_{i}}\ [(e_{\mu}^{L}e_{\nu}^{T_{i}}e_{\lambda}^{T_{i}%
}+\text{per.})-e_{\mu}^{L}e_{\nu}^{L}e_{\lambda}^{L}]\nonumber\\
&  +\sum_{(i\neq j)}u_{LT_{i}T_{j}}\ [e_{\mu}^{L}e_{\nu}^{T_{i}}e_{\lambda
}^{T_{j}}+\text{per.}]\nonumber\\
&  +\sum_{i}u_{T_{i}T_{i}T_{i}}\ [e_{\mu}^{T_{i}}e_{\nu}^{T_{i}}e_{\lambda
}^{T_{i}}-(e_{\mu}^{L}e_{\nu}^{L}e_{\lambda}^{T_{i}}+\text{per.})]\nonumber\\
&  +\sum_{i\neq j}u_{T_{j}T_{j}T_{i}}\ [(e_{\mu}^{T_{j}}e_{\nu}^{T_{j}%
}e_{\lambda}^{T_{i}}+\text{per.})-(e_{\mu}^{L}e_{\nu}^{L}e_{\lambda}^{T_{i}%
}+\text{per.})]\nonumber\\
&  +\sum_{(i\neq j\neq k)}u_{T_{i}T_{j}T_{k}}\ [e_{\mu}^{T_{i}}e_{\nu}^{T_{j}%
}e_{\lambda}^{T_{k}}+\text{per.}],\tag{2.60}%
\end{align}%
\begin{align}
\frac{1}{3}\varepsilon_{(\mu\nu)} &  =u_{PPP}(e_{\mu}^{P}e_{\nu}^{P}+3e_{\mu
}^{L}e_{\nu}^{L})\nonumber\\
&  +u_{PPL}(e_{\mu}^{P}e_{\nu}^{L}+e_{\mu}^{L}e_{\nu}^{P})\nonumber\\
&  +\sum_{i}u_{PPT_{i}}(e_{\mu}^{P}e_{\nu}^{T_{i}}+e_{\mu}^{T_{i}}e_{\nu}%
^{P})\nonumber\\
&  +\sum_{i}u_{PLT_{i}}(e_{\mu}^{L}e_{\nu}^{T_{i}}+e_{\mu}^{T_{i}}e_{\nu}%
^{L})\nonumber\\
&  +\sum_{i}u_{PT_{i}T_{i}}(e_{\mu}^{T_{i}}e_{\nu}^{T_{i}}-e_{\mu}^{L}e_{\nu
}^{L})\nonumber\\
&  +\sum_{(i\neq j)}u_{PT_{i}T_{j}}(e_{\mu}^{T_{i}}e_{\nu}^{T_{j}}+e_{\nu
}^{T_{j}}e_{\mu}^{T_{j}}),\tag{2.61}%
\end{align}

\begin{equation}
\frac{1}{2}\varepsilon_{\mu}=[u_{PPP}\ e_{\mu}^{P}+u_{PPL}\ e_{\mu}^{L}%
+\sum_{i}u_{PPT_{i}}\ e_{\mu}^{T_{i}}].\tag{2.62}%
\end{equation}
Putting all these polarizations back to the general form of physical states
Eq.(2.53), we get
\begin{align}
&  [\varepsilon_{\mu\nu\lambda}\alpha_{-1}^{\mu}\alpha_{-1}^{\nu}\alpha
_{-1}^{\lambda}+\varepsilon_{(\mu\nu)}\alpha_{-1}^{\mu}\alpha_{-2}^{\nu
}+\varepsilon_{\mu}\alpha_{-3}^{\mu}]|0,k\rangle\nonumber\\
&  =|A(\tilde{u})\rangle+|C(\theta)\rangle\nonumber\\
&  +[\frac{1}{9}(u_{LLL}+\sum_{i}u_{T_{i}T_{i}L})]|D_{1}(e^{L})\rangle
\nonumber\\
&  +\sum_{i}[\frac{1}{9}(u_{LLT_{i}}+\sum_{j}u_{T_{j}T_{j}T_{i}}%
)]|D_{1}(e^{T_{i}})\rangle\nonumber\\
&  +\frac{1}{75}[u_{LLP}+\sum_{i}u_{PT_{i}T_{i}}]|E\rangle.\tag{2.63}%
\end{align}
For the first two terms on the right hand side of Eq.(2.63), we need to make
the following replacements. For the positive-norm state $|A(\tilde{u})\rangle$
in Eq.(2.38)
\begin{align}
&  \tilde{u}_{T_{i}T_{i}T_{i}}\rightarrow u_{T_{i}T_{i}T_{i}}-\frac{1}%
{3}u_{PPT_{i}},\hspace{1cm}\tilde{u}_{T_{j}T_{j}T_{i}}\rightarrow
u_{T_{j}T_{j}T_{i}}-\frac{1}{9}u_{PPT_{i}},\nonumber\\
&  \tilde{u}_{T_{i}T_{j}T_{k}}\rightarrow u_{T_{i}T_{j}T_{k}}\hspace
{0.7cm}\tilde{u}_{LT_{i}T_{i}}\rightarrow u_{LT_{i}T_{i}}-\frac{1}{9}%
u_{PPL},\hspace{0.7cm}\tilde{u}_{LT_{i}T_{j}}\rightarrow u_{LT_{i}T_{j}%
}.\tag{2.64}%
\end{align}
For the spin-two zero-norm state $|C(\theta)\rangle$ in Eq.(2.43), the
replacement is given by
\begin{equation}
2\theta_{LT_{i}}\rightarrow3u_{PLT_{i}},\hspace{1cm}2\theta_{T_{i}T_{j}%
}\rightarrow3u_{PT_{i}T_{j}},\ \text{for }i\neq j,\hspace{1cm}2\theta
_{T_{i}T_{i}}\rightarrow3(u_{PT_{i}T_{i}}-\frac{3}{25}u_{PPP}).\tag{2.65}%
\end{equation}

It is important to note that for the spin-three gauge multiplet, only
spin-two, singlet and $D_{1}$ vector zero-norm states appear in the
decomposition Eq.(2.63). In the next subsection, we will see how one can
include the missing $D_{2}$ zero-norm state in the analysis.

\subsubsection{Complete spectrum analysis and the $D_{2}$ zero-norm state}

After all these preparations, we are ready for a complete analysis of the most
general decomposition of physical states at $m^{2}=4$. The most general form
of physical states at this mass level are given by
\begin{equation}
\lbrack\epsilon_{\mu\nu\lambda}\alpha_{-1}^{\mu}\alpha_{-1}^{\nu}\alpha
_{-1}^{\lambda}+\epsilon_{(\mu\nu)}\alpha_{-1}^{\mu}\alpha_{-2}^{\nu}%
+\epsilon_{\lbrack\mu\nu]}\alpha_{-1}^{\mu}\alpha_{-2}^{\nu}+\epsilon_{\mu
}\alpha_{-3}^{\mu}]|0,k\rangle. \tag{2.66}%
\end{equation}
The Virasoro constraints are
\begin{align}
\epsilon_{(\mu\nu)}  &  =-\frac{3}{2}k^{\lambda}\epsilon_{\mu\nu\lambda
},\tag{2.67}\\
-k^{\nu}\epsilon_{\lbrack\mu\nu]}+3\epsilon_{\mu}  &  =\frac{3}{2}k^{\nu
}k^{\lambda}\epsilon_{\mu\nu\lambda},\tag{2.68}\\
2k^{\nu}\epsilon_{\lbrack\mu\nu]}+3\epsilon_{\mu}  &  =3(k^{\nu}k^{\lambda
}-\eta^{\nu\lambda})\epsilon_{\mu\nu\lambda}. \tag{2.69}%
\end{align}
The solutions to Eqs.(2.68) and (2.69) are given by
\begin{align}
k^{\nu}\epsilon_{\lbrack\mu\nu]}  &  =(\frac{1}{2}k^{\nu}k^{\lambda}-\eta
^{\nu\lambda})\epsilon_{\mu\nu\lambda},\tag{2.70}\\
3\epsilon_{\mu}  &  =(2k^{\nu}k^{\lambda}-\eta^{\nu\lambda})\epsilon_{\mu
\nu\lambda}. \tag{2.71}%
\end{align}

In contrast to the previous discussion Eqs.(2.54),(2.55), where both
$\epsilon_{(\mu\nu)}$ and $\epsilon_{\mu}$ are completely fixed by the leading
spin-three polarization tensor $\epsilon_{\mu\nu\lambda}$, we now have a new
contribution from $k^{\nu}\epsilon_{\lbrack\mu\nu]}$. It will become clear
that this extra term includes the inter-particle zero-norm state $D_{2}$,
Eqs.(2.49) or (2.50). Furthermore, it should be clear that the antisymmetric
spin-two positive-norm physical states are defined by requiring $\epsilon
_{\mu\nu\lambda}=\epsilon_{(\mu\nu)}=0$ and $\epsilon_{\mu}=k^{\nu}%
\epsilon_{\lbrack\mu\nu]}=0$. In the following, for the sake of clarity, we
shall focus on the effects of the new contribution induced by the
$\epsilon_{\lbrack\mu\nu]}$ only.

The two independent polarization tensors of the most general representation
for physical states Eq.(2.66) are given in the helicity basis by
\begin{align}
\epsilon_{\mu\nu\lambda} &  =\sum_{ABC}U_{ABC}\ e_{\mu}^{A}e_{\nu}%
^{B}e_{\lambda}^{C},\hspace{1cm}A,B,C=P,L,T_{i};\tag{2.72}\\
\epsilon_{\lbrack\mu\nu]} &  =\sum_{A,B}V_{[AB]}\ e_{\mu}^{A}e_{\nu}%
^{B}.\tag{2.73}%
\end{align}
The Virasoro constraint Eq.(2.70) demands that
\begin{align}
3U_{PPP}-U_{LLP}-\sum_{i}U_{PT_{i}T_{i}} &  =0,\tag{2.74}\\
3U_{PPL}-U_{LLL}-\sum_{i}U_{LT_{i}T_{i}} &  =2V_{[PL]},\tag{2.75}\\
3U_{PPT_{i}}-U_{LLT_{i}}-\sum_{j}U_{T_{j}T_{j}T_{i}} &  =2V_{[PT_{i}%
]}.\tag{2.76}%
\end{align}
In contrast to Eq.(2.59), the solution to the above equations become
\begin{align}
U_{PPL}=U_{PPL}^{(1)}+U_{PPL}^{(2)}, &  \text{ \ where \ }U_{PPL}^{(1)}%
=\frac{1}{3}(U_{LLL}+\sum_{i}U_{T_{i}T_{i}L}),\text{ }U_{PPL}^{(2)}=\frac
{2}{3}V_{[PL]}\text{ ;}\tag{2.77}\\
U_{PPT_{i}}=U_{PPT_{i}}^{(1)}+U_{PPT_{i}}^{(2)}, &  \text{ \ where
\ \ }U_{PPT_{i}}^{(1)}=\frac{1}{3}(U_{LLT_{i}}+\sum_{j}U_{T_{j}T_{j}T_{i}%
}),\text{ }U_{PPT_{i}}^{(2)}=\frac{2}{3}V_{[PT_{i}]}.\tag{2.78}%
\end{align}
It is clear from the expressions above that only $U_{PPL}^{(2)}$ and
$U_{PPT_{i}}^{(2)}$ give new contributions to our previous analysis in the
last subsection, so we can simply write down all these new terms
\begin{align}
\delta\epsilon_{\mu\nu\lambda} &  =\frac{2}{3}[V_{[PL]}(e_{\mu}^{P}e_{\nu}%
^{P}e_{\lambda}^{L}+\text{per.})+\sum_{i}V_{[PT_{i}]}(e_{\mu}^{P}e_{\nu}%
^{P}e_{\lambda}^{T_{i}}+\text{per.})],\tag{2.79}\\
\delta\epsilon_{\lbrack\mu\nu]} &  =V_{[PL]}(e_{\mu}^{P}e_{\nu}^{L}%
-\text{per.})+\sum_{i}V_{[PT_{i}]}(e_{\mu}^{P}e_{\nu}^{T_{i}}-\text{per.}%
)\nonumber\\
&  +\sum_{i}V_{[T_{i}L]}(e_{\mu}^{T_{i}}e_{\nu}^{L}-\text{per.})+\sum_{i\neq
j}V_{[T_{j}T_{i}]}(e_{\mu}^{T_{j}}e_{\nu}^{T_{i}}-\text{per.}),\tag{2.80}\\
\delta\epsilon_{(\mu\nu)} &  =2[V_{[PL]}(e_{\mu}^{P}e_{\nu}^{L}+\text{per.}%
)+\sum_{i}V_{[PT_{i}]}(e_{\mu}^{P}e_{\nu}^{T_{i}}+\text{per.})],\tag{2.81}\\
\delta\epsilon_{\mu} &  =2[V_{[PL]}e_{\mu}^{L}+\sum_{i}V_{[PT_{i}]}e_{\mu
}^{T_{i}}].\tag{2.82}%
\end{align}
Finally, the complete decomposition of physical states Eq.(2.66) in the
helicity basis becomes
\begin{align}
&  [\epsilon_{\mu\nu\lambda}\alpha_{-1}^{\mu}\alpha_{-1}^{\nu}\alpha
_{-1}^{\lambda}+\epsilon_{(\mu\nu)}\alpha_{-1}^{\mu}\alpha_{-2}^{\nu}%
+\epsilon_{\lbrack\mu\nu]}\alpha_{-1}^{\mu}\alpha_{-2}^{\nu}+\epsilon_{\mu
}\alpha_{-3}^{\mu}]|0,k\rangle\nonumber\\
&  =|A(U_{CBA})\rangle+|B(V_{[T_{i}A]})\rangle+|C(U_{PBA})\rangle\tag{2.83}\\
&  +\sum_{A=L,T_{i}}[\frac{1}{9}(U_{LLA}+\sum_{i}U_{T_{i}T_{i}A})]|D_{1}%
(e^{A})\rangle\tag{2.84}\\
&  -\frac{1}{9}\sum_{A=L,T_{i}}V_{[PA]}|D_{2}^{\prime}(e^{A})\rangle
\tag{2.85}\\
&  +\frac{1}{75}[U_{LLP}+\sum_{i}U_{PT_{i}T_{i}}]|E\rangle.\tag{2.86}%
\end{align}
In Eq.(2.83), $|A(U_{CBA})\rangle$ is given by Eq.(2.38) with $\tilde{u}%
_{CBA}$ given by Eq.(2.64) and we have replaced $u$ by $U$ on the r.h.s. of
Eq.(2.64). The antisymmetric spin-two positive-norm state $|B(V_{[T_{i}%
A]})\rangle$ is given by Eq.(2.42) and we have replaced $v$ by $V$ \ in Eq.
(2.42). Finally, $|C(U_{PBA})\rangle$ is given by Eq.(2.43) with $\theta$
given by Eq.(2.65) and we have replaced $u$ by $U$ on the r.h.s. of Eq.(2.65).
In Eq.(2.85), $|D_{2}^{\prime}(e^{A})\rangle\equiv|D_{2}(e^{A})\rangle
-2|D_{1}(e^{A})\rangle$ is the inter-particle zero-norm state introduced in
the end of subsection A with $\alpha=-2$. Note that the value of $\alpha$ is a
choice of convension fixed by the parametrizations of the polarizations. It
can always be adjusted to be zero. In view of Eqs.(2.77) and (2.78), we see
that \textit{one can use either }$V_{[PA]}$\textit{ or }$U_{PPA}^{(2)}%
$\textit{ ( }$A=L,T_{i})$\textit{ to represent the polarization of the
}$|D_{2}^{\prime}(e^{A})\rangle$ \textit{inter-particle zero-norm state.}

We conclude that once we turn on the antisymmetric spin-two positive-norm
state in the general representation of physical states Eq.(2.66), it is
naturally accompanied by the $D_{2}^{\prime}$ inter-particle zero-norm state.
The polarization of the $D_{2}^{\prime}$ inter-particle zero-norm state can be
represented by either $V_{[PA]}$ or $U_{PPA}^{(2)}$ ( $A=L,T_{i})$ in
Eqs.(2.72) and (2.73). Thus this inter-particle zero-norm state will generate
an inter-particle symmetry transformation in the $\sigma$-model calculation
considered in \cite{7,13}. Note that, in contrast to the high-energy symmetry
of Gross \cite{5}, this symmetry is valid to all orders in $\alpha^{\prime}$.

\section{\bigskip Light-cone DDF Zero-norm States}

In the usual light-cone quantization of bosonic string theory, one solves the
Virasoro constraints to get rid of two string coordinates $X^{\pm}$. Only 24
string coordinates $\alpha_{n}^{i},$ $i=1,...,,24,$ remain, and there are no
zero-norm states in the spectrum. However, there exists another related
quantization scheme, the DDF quantization, which does include the zero-norm
states in the spectrum. In the light-cone DDF quantization of open bosonic
string \cite{17}, one constructs transverse physical states with discrete momenta%

\begin{equation}
\ p^{\mu}=\ p_{0}^{\mu}-Nk_{0}^{\mu}=(1,0.....,-1+N),\tag{3.1}%
\end{equation}
where $X^{\pm}$ $\equiv\frac{1}{\sqrt{2}}(X^{0}$ $\pm X^{25})$ and
$\ p^{+}=1,$ $p^{-}=-1+N.$ In Eq.(3.1) $m^{2}=-p^{2}=2(N-1)$ and $p_{0}^{\mu
}\equiv(1,0...,-1)$, $k_{0}^{\mu}\equiv(0,0...,-1)$, respectively. All other
momenta can be reached by Lorentz transformations. The DDF operators are given
by \cite{17}%
\begin{equation}
\ A_{n}^{i}=\frac{1}{2\pi}%
%TCIMACRO{\tint \limits_{0}^{2\pi}}%
%BeginExpansion
{\textstyle\int\limits_{0}^{2\pi}}
%EndExpansion
\dot{X}^{i}(\tau)e^{inX^{+}(\tau)}d\tau,\text{ }i=1,...,,24,\tag{3.2}%
\end{equation}
where the massless vertex operator $V^{i}(nk_{0},\tau)=\dot{X}^{i}%
(\tau)e^{inX^{+}(\tau)}$ is a primary field with conformal dimension one, and
is periodic in the worldsheet time $\tau$ if one chooses $k^{\mu}=$
$nk_{0}^{\mu}$ with $n\in Z.$ It is then easy to show that%
\begin{align}
\ [L_{m},A_{n}^{i}] &  =0,\tag{3.3}\\
\lbrack A_{m}^{i},A_{n}^{j}] &  =m\delta_{ij}\delta_{m+n}.\tag{3.4}%
\end{align}
In addition to sharing the same algebra, Eq.(3.4), with string coordinates
$\alpha_{n}^{i},$ the DDF operators $A_{n}^{i}$ possess a nicer property
Eq.(3.3), which enables us to easily write down a general formula for the
positive-norm physical states as following%
\begin{equation}
\ (A_{-1}^{j})^{i_{1}}(A_{-2}^{k})^{i_{2}}....(A_{-m}^{l})^{i_{m}}\mid
0,p_{0}>,\text{ }i_{r}\in N,\tag{3.5}%
\end{equation}
where $\mid0,p_{0}>$ is the tachyon ground state and $N=%
%TCIMACRO{\tsum _{r=1}^{m}}%
%BeginExpansion
{\textstyle\sum_{r=1}^{m}}
%EndExpansion
ri_{r}$ is the level of the state. Historically, DDF operators were used to
prove no-ghost (negative-norm states) theorem for $D=26$ string theory. Here
we are going to use them to analyse zero-norm states. It turns out that
zero-norm states can be generated by%
\begin{equation}
\ \tilde{A}_{n}^{-}=A_{n}^{-}-\sum_{m=1}^{\infty}\sum_{i=1}^{D-2}:A_{m}%
^{i}A_{n-m}^{i}:,\tag{3.6}%
\end{equation}
where $A_{n}^{-}$ is given by%
\begin{equation}
\ A_{n}^{-}=\frac{1}{2\pi}%
%TCIMACRO{\tint \limits_{0}^{2\pi}}%
%BeginExpansion
{\textstyle\int\limits_{0}^{2\pi}}
%EndExpansion
[:\dot{X}^{-}e^{inX^{+}}:-\frac{1}{2}in\frac{d}{d\tau}(\log\dot{X}%
^{+})e^{inX^{+}}]d\tau.\tag{3.7}%
\end{equation}
It can be shown that $\tilde{A}_{n}^{-}$ commute with $L_{m}$ and satisfy the
following algebra%
\begin{align}
\ [\tilde{A}_{m}^{-},A_{n}^{i}] &  =0,\tag{3.8}\\
\lbrack\tilde{A}_{m}^{-},\tilde{A}_{n}^{-}] &  =(m-n)\tilde{A}_{m+n}^{-}%
+\frac{26-D}{12}m^{3}\delta_{m+n}.\tag{3.9}%
\end{align}
Eqs.(3.4), (3.8) and (3.9) constitute the spectrum generating algebra for the
open bosonic string including zero-norm states. The ground state $\mid
0,p_{0}>\equiv\mid0>$ satisfies the following conditions%
\begin{align}
A_{n}^{i} &  \mid0>=\ \tilde{A}_{n}^{-}\mid0>=0,\text{ \ }n>0,\tag{3.10}\\
\ \tilde{A}_{0}^{-} &  \mid0>=-\frac{26-D}{24},\text{ \ }A_{0}^{i}%
\mid0>=0.\tag{3.11}%
\end{align}
We are now ready to construct zero-norm states in the DDF formalism.

1. $m^{2}=0:$ One has only one scalar $\tilde{A}_{-1}^{-}\mid0>$, which has
zero-norm for any $D$.

2. $m^{2}=2:$ One has a light-cone vector $A_{-1}^{i}\tilde{A}_{-1}^{-}\mid
0>$, which has zero-norm for any $D$, and two scalars, whose norms are
calculated to be%
\begin{equation}
\parallel(\ a\tilde{A}_{-1}^{-}\tilde{A}_{-1}^{-}+b\tilde{A}_{-2}^{-}%
)\mid0>\parallel=\frac{26-D}{2}b^{2}.\tag{3.12}%
\end{equation}
For $b=0,$ one has a "pure type I" zero-norm state, $\tilde{A}_{-1}^{-}%
\tilde{A}_{-1}^{-}\mid0>$, which has zero-norm for any $D$. By combining with
the light-cone vector $A_{-1}^{i}\tilde{A}_{-1}^{-}\mid0>,$ one obtains a
vector zero-norm state with 25 degrees of freedom, which correspond to
Eq.(2.5) in the OCFQ approach. For $b\neq0$, one obtains a type II scalar
zero-norm state for $D=26,$ which corresponds to Eq.(2.4) in the OCFQ approach.

3. $m^{2}=4:$

I. A spin-two tensor $A_{-1}^{i}A_{-1}^{j}\tilde{A}_{-1}^{-}\mid0>$, which has
zero-norm for any $D$.

II. Three light-cone vectors, whose norms are calculated to be%
\begin{equation}
\parallel(\ aA_{-1}^{i}\tilde{A}_{-1}^{-}\tilde{A}_{-1}^{-}+bA_{-2}^{i}%
\tilde{A}_{-1}^{-}+cA_{-1}^{i}\tilde{A}_{-2}^{-})\mid0>\parallel=\frac
{26-D}{2}c^{2}. \tag{3.13}%
\end{equation}
\qquad\qquad

III. Three scalars, whose norms are calculated to be%
\begin{equation}
\parallel(\ d\tilde{A}_{-1}^{-}\tilde{A}_{-1}^{-}\tilde{A}_{-1}^{-}+e\tilde
{A}_{-1}^{-}\tilde{A}_{-2}^{-}+f\tilde{A}_{-3}^{-})\mid0>\parallel
=2(26-D)(e+f)^{2}.\tag{3.14}%
\end{equation}
For $c=0$ in Eq.(3.13), one has two "pure type I" light-cone vector zero-norm
states. For $e+f=0$ in Eq.(3.14), one has two "pure type I" scalar zero-norm
states. One of the two type I light-cone vectors, when combining with the
spin-two state in I, gives the type I spin-two tensor which corresponds to
Eq.(2.7) in the OCFQ approach. The other type I light-cone vector, when
combining with one of the two type I scalar, gives the type I vector zero-norm
state which corresponds to Eq.(2.8) in the OCFQ approach. The other type I
scalar corresponds to Eq.(2.9). Finally, for $c\neq0$ and $e+f\neq0$, one
obtains the type II vector zero-norm state for $D=26$, which corresponds to
Eq.(2.6) in the OCFQ approach. It is easy to see that a special linear
combination of $b$ and $c$ will give the inter-particle vector zero-norm state
which corresponds to the inter-particle $D_{2}$ zero-norm state in Eq (2.10).
This completes the analysis of zero-norm states for $m^{2}=4.$ Note that the
exact mapping of zero-norm states in the light-cone DDF formalism and the OCFQ
approach depends on the exact relation between operators $(\tilde{A}_{n}%
^{-},A_{n}^{i},L_{n})$ and $\alpha_{n}^{\mu}$, which has not been worked out
in the literature.

\section{BRST Zero-norm States in WSFT}

Cubic string field theory is defined on a disk with the action
\begin{equation}
S=-\frac{1}{g_{0}}\left(  \frac{1}{2}\int\Phi\ast Q_{\text{B}}\Phi+\frac{1}%
{3}\int\Phi\ast\Phi\ast\Phi\right)  ,\tag{4.1}%
\end{equation}
where $Q_{\text{B}}$ is the BRST charge%
\begin{equation}
Q_{\text{B}}=\sum_{n=-\infty}^{\infty}L_{-n}^{\text{m}}c_{n}+\sum
_{m,n=-\infty}^{\infty}\dfrac{m-n}{2}:c_{m}c_{n}b_{-m-n}:-c_{0},\tag{4.2}%
\end{equation}
and $\Phi$ is the string field with ghost number $1$ and $b,c$ are conformal
ghosts. Since the ghost number of vacuum on a disk is -3, the total ghost
number of this action is 0 as expected. The string field can be expanded as
\[
\Phi=\sum_{k,m,n}A_{\mu\cdots,k\cdots m\cdots n\cdots}\left(  x\right)
\alpha_{k}^{\mu}\cdots b_{m}\cdots c_{n}\cdots\left\vert \Omega\right\rangle ,
\]
where the string ground state $\left\vert \Omega\right\rangle $\ is
\begin{equation}
\left\vert \Omega\right\rangle =c_{1}\left\vert 0\right\rangle .\tag{4.3}%
\end{equation}
The gauge transformation for string field can be written as
\begin{equation}
\delta\Phi=Q_{\text{B}}\Lambda+g\left(  \Phi\ast\Lambda-\Lambda\ast
\Phi\right)  ,\tag{4.4}%
\end{equation}
where $\Lambda$ is the a string field with ghost number $0$.

For the purpose of discussion in this paper, we are going to consider the
linearized gauge transformation
\begin{equation}
\delta\Phi=Q_{\text{B}}\Lambda,\tag{4.5}%
\end{equation}
where $Q_{\text{B}}\Lambda$ is just the off-shell zero-norm states. In the
following, we will explicitly show that the components of Eq.(4.5) are in
one-to-one correspondence to the zero-norm states obtained in OCFQ approach in
section II level by level for the first few mass levels.

There is no zero-norm state in the lowest string mass level with $m^{2}=-2$,
so our analysis will start with the mass level of $m^{2}=0$.

\bigskip

\noindent\underline{$m^{2}=0$:}

The string field can be expanded as%
\begin{align}
\Phi &  =\left\{  iA_{\mu}\left(  x\right)  \alpha_{-1}^{\mu}+\alpha\left(
x\right)  b_{-1}c_{0}\right\}  \left\vert \Omega\right\rangle ,\tag{4.6}\\
\Lambda &  =\left\{  \epsilon^{0}\left(  x\right)  b_{-1}\right\}  \left\vert
\Omega\right\rangle .\tag{4.7}%
\end{align}
The gauge transformation is then%
\begin{equation}
Q_{\text{B}}\Lambda=\left\{  -\dfrac{1}{2}\alpha_{0}^{2}\epsilon^{0}%
b_{-1}c_{0}+\epsilon^{0}\alpha_{0}\cdot\alpha_{-1}\right\}  \left\vert
\Omega\right\rangle .\tag{4.8}%
\end{equation}
The nilpotency of BRST charge $Q_{\text{B}}$ gives%
\begin{equation}
Q_{\text{B}}^{2}\Lambda=0,\tag{4.9}%
\end{equation}
which can be easily checked to be valid for any $D$. Thus Eq.(4.8) can be
interpreted as a type I zero-norm state. To compare it with the zero-norm
state obtained in OCFQ approach in section II, we need to reduce the Hilbert
space by removing the ghosts states. In particular, the coefficients of terms
with ghost operaters must vanish. For the state in Eq.(4.8), it is%
\begin{equation}
\alpha_{0}^{2}\epsilon^{0}=0,\tag{4.10}%
\end{equation}
which give the on-shell condition $k^{2}=0$ and the following zero-norm state%
\begin{equation}
Q_{\text{B}}\Lambda=\epsilon^{0}\alpha_{0}\cdot\alpha_{-1}\left\vert
\Omega\right\rangle .\tag{4.11}%
\end{equation}
This is the same as the scalar zero-norm state obtained in OCFQ approach.

\bigskip

\noindent\underline{$m^{2}=2$:}

The string fields expansion are%
\begin{align}
\Phi &  =\left\{  -B_{\mu\nu}\left(  x\right)  \alpha_{-1}^{\mu}\alpha
_{-1}^{\nu}+iB_{\mu}\left(  x\right)  \alpha_{-2}^{\mu}\right.  \nonumber\\
&  \text{ \ \ \ \ \ }\left.  +i\beta_{\mu}\left(  x\right)  \alpha_{-1}^{\mu
}b_{-1}c_{0}+\beta^{0}\left(  x\right)  b_{-2}c_{0}+\beta^{1}\left(  x\right)
b_{-1}c_{-1}\right\}  \left\vert \Omega\right\rangle ,\tag{4.12}\\
\Lambda &  =\left\{  i\epsilon_{\mu}^{0}\left(  x\right)  \alpha_{-1}^{\mu
}b_{-1}+\epsilon^{1}\left(  x\right)  b_{-2}\right\}  \left\vert
\Omega\right\rangle .\tag{4.13}%
\end{align}
The off-shell zero-norm states are calculated to be%
\begin{align}
Q_{\text{B}}\Lambda &  =\left\{  \left(  i\alpha_{0\mu}\epsilon_{\nu}%
^{0}+\frac{1}{2}\epsilon^{1}\eta_{\mu\nu}\right)  \alpha_{-1}^{\mu}\alpha
_{-1}^{\nu}+\left(  i\epsilon^{0}+\alpha_{0}\epsilon^{1}\right)  \cdot
\alpha_{-2}\right.  \nonumber\\
&  \text{ \ \ \ \ \ }-i\dfrac{1}{2}\left(  \alpha_{0}^{2}+2\right)  \left(
\epsilon^{0}\cdot\alpha_{-1}\right)  b_{-1}c_{0}-\dfrac{1}{2}\left(
\alpha_{0}^{2}+2\right)  \epsilon^{1}b_{-2}c_{0}\nonumber\\
&  \text{ \ \ \ \ \ }\left.  -\left(  i\alpha_{0}\cdot\epsilon^{0}%
+3\epsilon^{1}\right)  b_{-1}c_{-1}\right\}  \left\vert \Omega\right\rangle
.\tag{4.14}%
\end{align}
Nilpotency condition requires%
\begin{equation}
Q_{\text{B}}^{2}\Lambda=\dfrac{D-26}{2}\epsilon^{1}c_{-2}\left\vert
\Omega\right\rangle =0.\tag{4.15}%
\end{equation}
There are two solutions of Eq.(4.15), which correspond to the type I and type
II zero-norm states, respectively.

\begin{enumerate}
\item Type I: in this case $D$ is not restricted to the critical string
dimension in Eq.(4.15), i.e. $D\neq26$. Thus
\begin{equation}
\epsilon^{1}=0.\tag{4.16}%
\end{equation}
The no-ghost conditions of Eq.(4.14) lead to the on-shell constraints%
\begin{align}
\alpha_{0}^{2}+2 &  =0,\tag{4.17}\\
\alpha_{0}\cdot\epsilon^{0} &  =0.\tag{4.18}%
\end{align}
The off-shell zero-norm state in Eq.(4.14) then reduces to an on-shell vector
zero-norm state%
\begin{equation}
Q_{\text{B}}\Lambda=i\left\{  \left(  \epsilon^{0}\cdot\alpha_{-1}\right)
\left(  \alpha_{0}\cdot\alpha_{-1}\right)  +\epsilon^{0}\cdot\alpha
_{-2}\right\}  \left\vert \Omega\right\rangle \tag{4.19}%
\end{equation}

\item Type II: in this case $D$ is restricted to the critical string
dimension, i.e. $D=26$, and  $\epsilon^{1}$ can be arbitrary function. The
no-ghost conditions then lead to the on-shell constraints%
\begin{align}
\alpha_{0}^{2}+2 &  =0,\tag{4.20}\\
i\alpha_{0}\cdot\epsilon^{0}+3\epsilon^{1} &  =0.\tag{4.21}%
\end{align}
The second condition can be solved by a special solution%
\begin{equation}
\epsilon_{\mu}^{0}=-\dfrac{3i}{2}\alpha_{0\mu}\epsilon^{1},\tag{4.22}%
\end{equation}
which leads to an on-shell scalar zero-norm state%
\begin{equation}
Q_{\text{B}}\Lambda=\left\{  \dfrac{3}{2}\left(  \alpha_{0}\cdot\alpha
_{-1}\right)  ^{2}+\frac{1}{2}\left(  \alpha_{-1}\cdot\alpha_{-1}\right)
+\dfrac{5}{2}\left(  \alpha_{0}\cdot\alpha_{-2}\right)  \right\}  \epsilon
^{1}\left\vert \Omega\right\rangle \tag{4.23}%
\end{equation}

\end{enumerate}

Again, up to a constant factor, the zero-norm states Eqs.(4.19) and (4.23) are
the same as Eqs. (2.5) and (2.4) calculated in the OCFQ approach.

\noindent\underline{$m^{2}=4$:}

The string fields are expanded as%
\begin{align}
\Phi &  =\left\{  -iC_{\mu\nu\lambda}\left(  x\right)  \alpha_{-1}^{\mu}%
\alpha_{-1}^{\nu}\alpha_{-1}^{\lambda}-C_{\mu\nu}\left(  x\right)  \alpha
_{-2}^{\mu}\alpha_{-1}^{\nu}+iC_{\mu}\left(  x\right)  \alpha_{-3}^{\mu
}\right. \nonumber\\
&  \text{ \ \ \ \ \ }-\gamma_{\mu\nu}\left(  x\right)  \alpha_{-1}^{\mu}%
\alpha_{-1}^{\nu}b_{-1}c_{0}+i\gamma_{\mu}^{0}\left(  x\right)  \alpha
_{-1}^{\mu}b_{-2}c_{0}+i\gamma_{\mu}^{1}\left(  x\right)  \alpha_{-1}^{\mu
}b_{-1}c_{-1}\nonumber\\
&  \text{ \ \ \ \ \ }\left.  +i\gamma_{\mu}^{2}\left(  x\right)  \alpha
_{-2}^{\mu}b_{-1}c_{0}+\gamma^{0}\left(  x\right)  b_{-3}c_{0}+\gamma
^{1}\left(  x\right)  b_{-2}c_{-1}+\gamma^{2}\left(  x\right)  b_{-1}%
c_{-2}\right\}  \left\vert \Omega\right\rangle ,\tag{4.24}\\
\Lambda &  =\left\{  -\epsilon_{\mu\nu}\left(  x\right)  \alpha_{-1}^{\mu
}\alpha_{-1}^{\nu}b_{-1}+i\epsilon_{\mu}^{1}\left(  x\right)  \alpha_{-2}%
^{\mu}b_{-1}\right.  \left\vert \Omega\right\rangle \nonumber\\
&  \text{ \ \ \ \ \ }\left.  +i\epsilon_{\mu}^{2}\left(  x\right)  \alpha
_{-1}^{\mu}b_{-2}+\epsilon^{2}\left(  x\right)  b_{-3}+\epsilon^{3}\left(
x\right)  b_{-1}b_{-2}c_{0}\right\}  \left\vert \Omega\right\rangle .
\tag{4.25}%
\end{align}
The off-shell zero-norm states are%
\begin{align}
Q_{\text{B}}\Lambda &  =\left\{  \left(  -\alpha_{0\left(  \mu\right.
}\epsilon_{\left.  \nu\lambda\right)  }+\dfrac{i}{2}\epsilon_{\left(
\mu\right.  }^{2}\eta_{\left.  \nu\lambda\right)  }\right)  \alpha_{-1}^{\mu
}\alpha_{-1}^{\nu}\alpha_{-1}^{\lambda}+\left(  i\alpha_{0\mu}\epsilon_{\nu
}^{2}+i\alpha_{0\nu}\epsilon_{\mu}^{1}-2\epsilon_{\mu\nu}+\epsilon^{2}%
\eta_{\mu\nu}\right)  \alpha_{-2}^{\mu}\alpha_{-1}^{\nu}\right. \nonumber\\
&  \text{ \ \ \ \ }+\left(  \alpha_{0\mu}\epsilon^{2}+2i\epsilon_{\mu}%
^{1}+i\epsilon_{\mu}^{2}\right)  \alpha_{-3}^{\mu}+\left[  \dfrac{1}{2}\left(
\alpha_{0}^{2}+4\right)  \epsilon_{\mu\nu}+\dfrac{1}{2}\epsilon^{3}\eta
_{\mu\nu}\right]  \alpha_{-1}^{\mu}\alpha_{-1}^{\nu}b_{-1}c_{0}\nonumber\\
&  \text{ \ \ \ \ }+\left[  -\dfrac{i}{2}\left(  \alpha_{0}^{2}+4\right)
\epsilon_{\mu}^{2}-\alpha_{0\mu}\epsilon^{3}\right]  \alpha_{-1}^{\mu}%
b_{-2}c_{0}+\left(  2\alpha_{0}^{\nu}\epsilon_{\nu\mu}-2i\epsilon_{\mu}%
^{1}-3i\epsilon_{\mu}^{2}\right)  \alpha_{-1}^{\mu}b_{-1}c_{-1}\nonumber\\
&  \text{ \ \ \ \ }+\left[  -\dfrac{i}{2}\left(  \alpha_{0}^{2}+4\right)
\epsilon_{\mu}^{1}+\alpha_{0\mu}\epsilon^{3}\right]  \alpha_{-2}^{\mu}%
b_{-1}c_{0}+\left[  -\dfrac{1}{2}\left(  \alpha_{0}^{2}+4\right)  \epsilon
^{2}-\epsilon^{3}\right]  b_{-3}c_{0}\nonumber\\
&  \text{ \ \ \ \ }\left.  +\left(  -i\alpha_{0}^{\mu}\epsilon_{\mu}%
^{2}-4\epsilon^{2}-2\epsilon^{3}\right)  b_{-2}c_{-1}+\left(  -2i\alpha
_{0}^{\mu}\epsilon_{\mu}^{1}-5\epsilon^{2}+4\epsilon^{3}+\epsilon_{\mu}^{\mu
}\right)  b_{-1}c_{-2}\right\}  \left\vert \Omega\right\rangle . \tag{4.26}%
\end{align}
Nilpotency condition requires%
\begin{equation}
Q_{\text{B}}^{2}\Lambda=\left(  D-26\right)  \left[  \dfrac{i}{2}\epsilon
_{\mu}^{2}\alpha_{-1}^{\mu}c_{-2}+2\epsilon^{2}c_{-3}-\dfrac{1}{2}\epsilon
^{3}b_{-1}c_{-2}c_{0}\right]  =0. \tag{4.27}%
\end{equation}
Similarly, we classify the solutions of Eq. (4.27) by type I and type II in
the following:

\begin{enumerate}
\item Type I: $D\neq26$. This leads to%
\begin{equation}
\epsilon^{2}=\epsilon^{3}=\epsilon_{\mu}^{2}=0,\tag{4.28}%
\end{equation}
The no-ghost conditions lead to the on-shell constraints%
\begin{align}
\alpha_{0}^{2}+4 &  =0,\tag{4.29}\\
\alpha_{0}^{\nu}\epsilon_{\nu\mu}-i\epsilon_{\mu}^{1} &  =0,\tag{4.30}\\
-2i\left(  \alpha_{0}\cdot\epsilon^{1}\right)  +\epsilon_{\mu}^{\mu} &
=0.\tag{4.31}%
\end{align}
One can apply the same technique as in subsection IIB to obtain a complete set
of solutions to Eqs.(4.30) and (4.31). However, for simplicity, we shall list
all independent solutions only. There are three independent solutions to the
above equations, which correspond to the three type I on-shell zero-norm states:

\begin{itemize}
\item Tensor zero-norm state%
\begin{equation}
\epsilon_{\mu}^{1}=0\text{, \ \ }\alpha_{0}^{\nu}\epsilon_{\mu\nu}=0\text{,
\ \ }\epsilon_{\mu}^{\mu}=0, \tag{4.32}%
\end{equation}%
\begin{equation}
Q_{\text{B}}\Lambda=-\left\{  \alpha_{0\mu}\epsilon_{\nu\lambda}\alpha
_{-1}^{\mu}\alpha_{-1}^{\nu}\alpha_{-1}^{\lambda}+2\epsilon_{\mu\nu}%
\alpha_{-2}^{\mu}\alpha_{-1}^{\nu}\right\}  \left\vert \Omega\right\rangle .
\tag{4.33}%
\end{equation}

\item Vector zero-norm state%
\begin{equation}
\alpha_{0}\cdot\epsilon^{1}=0\text{, \ \ }\epsilon_{\mu\nu}=-\dfrac{i}%
{4}\left(  \alpha_{0\nu}\epsilon_{\mu}^{1}+\alpha_{0\mu}\epsilon_{\nu}%
^{1}\right)  ,\tag{4.34}%
\end{equation}%
\begin{align}
Q_{\text{B}}\Lambda &  =\left\{  i\dfrac{1}{2}\left(  \alpha_{0}\cdot
\alpha_{-1}\right)  ^{2}\left(  \epsilon^{1}\cdot\alpha_{-1}\right)
+2i\left(  \epsilon^{1}\cdot\alpha_{-3}\right)  \right.  \nonumber\\
&  \text{ \ \ \ \ \ }\left.  +\dfrac{3}{2}\left(  \alpha_{0}\cdot\alpha
_{-1}\right)  \left(  \epsilon^{1}\cdot\alpha_{-2}\right)  +\dfrac{1}%
{2}\left(  \alpha_{0}\cdot\alpha_{-2}\right)  \left(  \epsilon^{1}\cdot
\alpha_{-1}\right)  \right\}  \left\vert \Omega\right\rangle .\tag{4.35}%
\end{align}

\item Scalar zero-norm state%
\begin{equation}
\epsilon_{\mu}^{1}=\dfrac{i\left(  D-1\right)  }{9}\alpha_{0\mu}\theta\text{,
\ \ }\epsilon_{\mu\nu}=\eta_{\mu\nu}\theta+\dfrac{\left(  8+D\right)  }%
{36}\alpha_{0\mu}\alpha_{0\nu}\theta,\tag{4.36}%
\end{equation}%
\begin{align}
Q_{\text{B}}\Lambda &  =-\dfrac{2}{9}\left\{  \dfrac{\left(  8+D\right)  }%
{8}\left(  \alpha_{0}\cdot\alpha_{-1}\right)  ^{3}+\dfrac{9}{2}\left(
\alpha_{0}\cdot\alpha_{-1}\right)  \left(  \alpha_{-1}\cdot\alpha_{-1}\right)
+9\left(  \alpha_{-1}\cdot\alpha_{-2}\right)  \right.  \nonumber\\
&  \text{ \ \ \ }\left.  +\dfrac{3\left(  D+2\right)  }{4}\left(  \alpha
_{0}\cdot\alpha_{-1}\right)  \left(  \alpha_{0}\cdot\alpha_{-2}\right)
+\left(  D-1\right)  \left(  \alpha_{0}\cdot\alpha_{-3}\right)  \right\}
\theta\left\vert \Omega\right\rangle .\tag{4.37}%
\end{align}
If we set $D=26$, then%
\begin{align}
Q_{\text{B}}\Lambda &  =-\dfrac{2}{9}\left\{  \dfrac{17}{4}\left(  \alpha
_{0}\cdot\alpha_{-1}\right)  ^{3}+\dfrac{9}{2}\left(  \alpha_{0}\cdot
\alpha_{-1}\right)  \left(  \alpha_{-1}\cdot\alpha_{-1}\right)  +9\left(
\alpha_{-1}\cdot\alpha_{-2}\right)  \right.  \nonumber\\
&  \text{ \ \ \ }\left.  +21\left(  \alpha_{0}\cdot\alpha_{-1}\right)  \left(
\alpha_{0}\cdot\alpha_{-2}\right)  +25\left(  \alpha_{0}\cdot\alpha
_{-3}\right)  \right\}  \theta\left\vert \Omega\right\rangle ,\tag{4.38}%
\end{align}
where $\theta$ is an arbitrary function.
\end{itemize}

\item Type II: $D=26$ in Eq.(4.27), and $\epsilon^{2},\epsilon^{3}$ and
$\epsilon_{\mu}^{2}$ are arbitrary functions. The no-ghost conditions lead to
the on-shell constraints%
\begin{align}
\alpha_{0}^{2}+4 &  =0,\tag{4.39}\\
\epsilon^{3} &  =0,\tag{4.40}\\
2\alpha_{0}^{\nu}\epsilon_{\nu\mu}-2i\epsilon_{\mu}^{1}-3i\epsilon_{\mu}^{2}
&  =0,\tag{4.41}\\
i\alpha_{0}^{\mu}\epsilon_{\mu}^{2}+4\epsilon^{2} &  =0,\tag{4.42}\\
-2i\alpha_{0}^{\mu}\epsilon_{\mu}^{1}-5\epsilon^{2}+\epsilon_{\mu}^{\mu} &
=0.\tag{4.43}%
\end{align}
In addition to all three type I zero-norm states as found above, we now have a
new solution to the above equations. This special solution can be chosen as%
\begin{align}
\epsilon^{2} &  =-\dfrac{i}{4}\left(  \alpha_{0}\cdot\epsilon^{2}\right)
=0,\tag{4.44}\\
\epsilon_{\mu\nu} &  =-C\left(  \alpha_{0\mu}\epsilon_{\nu}^{2}+\alpha_{0\nu
}\epsilon_{\mu}^{2}\right)  ,\tag{4.45}\\
\epsilon_{\mu}^{1} &  =\dfrac{8iC-3}{2}\epsilon_{\mu}^{2},\tag{4.46}%
\end{align}
which gives an on-shell vector zero-norm state%
\begin{align}
Q_{\text{B}}\Lambda &  =i\left\{  \left(  8iC-2\right)  \left(  \epsilon
^{2}\cdot\alpha_{-3}\right)  +\dfrac{1}{2}\left(  \alpha_{-1}\cdot\alpha
_{-1}\right)  \left(  \epsilon^{2}\cdot\alpha_{-1}\right)  \right.
\nonumber\\
&  \text{ \ \ \ \ \ \ }+\left(  2iC+1\right)  \left(  \alpha_{0}\cdot
\alpha_{-2}\right)  \left(  \epsilon^{2}\cdot\alpha_{-1}\right)  +2iC\left(
\alpha_{0}\cdot\alpha_{-1}\right)  ^{2}\left(  \epsilon^{2}\cdot\alpha
_{-1}\right)  \nonumber\\
&  \text{ \ \ \ \ \ \ }\left.  +\dfrac{12iC-3}{2}\left(  \alpha_{0}\cdot
\alpha_{-1}\right)  \left(  \epsilon^{2}\cdot\alpha_{-2}\right)  \right\}
\left\vert \Omega\right\rangle .\tag{4.47}%
\end{align}
For a special value of $C=-3i/4$, Eq.(4.47) becomes%
\begin{align}
Q_{\text{B}}\Lambda &  =i\left\{  4\left(  \epsilon^{2}\cdot\alpha
_{-3}\right)  +\dfrac{1}{2}\left(  \alpha_{-1}\cdot\alpha_{-1}\right)  \left(
\epsilon^{2}\cdot\alpha_{-1}\right)  +\dfrac{5}{2}\left(  \alpha_{0}%
\cdot\alpha_{-2}\right)  \left(  \epsilon^{2}\cdot\alpha_{-1}\right)  \right.
\nonumber\\
&  \text{ \ \ \ \ \ \ }\left.  +\dfrac{3}{2}\left(  \alpha_{0}\cdot\alpha
_{-1}\right)  ^{2}\left(  \epsilon^{2}\cdot\alpha_{-1}\right)  +3\left(
\alpha_{0}\cdot\alpha_{-1}\right)  \left(  \epsilon^{2}\cdot\alpha
_{-2}\right)  \right\}  \left\vert \Omega\right\rangle .\tag{4.48}%
\end{align}
Up to a constant factor, zero-norm states in Eqs.(4.33), (4.35), (4.38) and
(4.48) are exactly the same as Eqs.( 2.7), (2.8), (2.9) and (2.6) calculated
in the OCFQ approach. In addition, it can be checked that for $C=-5i/8$ and
$-i/16$ in Eq.(4.47), one gets $D_{1}$ and $D_{2}$ zero-norm states of OCFQ
approach in Eqs.(11) and (10), respectively.

\qquad In Ref [16], the background ghost transformations in the gauge
transformations of WSFT \cite{15} were shown to correspond, in a one-to-one
manner, to the lifting of on-shell conditions of zero-norm states in the OCFQ
approach. For the rest of this section, we are going to go one step further
and apply the results calculated above to demonstrate that off-shell gauge
transformations of WSFT are indeed identical to the on-shell stringy gauge
symmetries generated by two types of zero-norm states in the generalized
massive $\sigma$-model approach \cite{7} of string theory. For the mass level
$m^{2}=2$, by using Eqs.(4.12) and (4.13), the linearized gauge transformation
of WSFT in Eq.(4.5) gives
\begin{align}
\delta B_{\mu\nu} &  =-\partial_{(\mu}\epsilon_{\nu)}^{0}-\frac{1}{2}%
\epsilon^{1}\eta_{\mu\nu},\tag{4.49}\\
\delta B_{\mu} &  =-\partial_{\mu}\epsilon^{1}+\frac{1}{2}\epsilon_{\mu}%
^{0},\tag{4.50}\\
\delta\beta_{\mu} &  =\frac{1}{2}(\partial^{2}-2)\epsilon_{\mu}^{0}%
,\tag{4.51}\\
\delta\beta^{0} &  =\frac{1}{2}(\partial^{2}-2)\epsilon^{1},\tag{4.52}\\
\delta\beta^{1} &  =-\partial^{\mu}\epsilon_{\mu}^{0}-3\epsilon^{1}.\tag{4.53}%
\end{align}
For the type I gauge transformation induced by zero-norm state in Eq.(2.5),
one can use Eqs.(4.16) -(4.18) to eliminate the background ghost
transformations Eqs.(4.51)-(4.53). Finally, conditions of worldsheet conformal
invariance in the presence of weak background fields \cite{7} can be used to
express $B_{\mu}$ in terms of $B_{\mu\nu}$, and one ends up with the following
on-shell gauge transformation by Eq.(4.49)%
\begin{equation}
\delta B_{\mu\nu}=\partial_{(\mu}\epsilon_{\nu)}^{0};\text{ \ \ }\partial
^{\mu}\epsilon_{\mu}^{0}=0,\text{ \ }(\partial^{2}-2)\epsilon_{\mu}%
^{0}=0.\tag{4.54}%
\end{equation}
Similarly, one can apply the same procedure to type II zero-norm state in
Eq.(2.4), and derive the following type II gauge transformation%
\begin{equation}
\delta B_{\mu\nu}=\frac{3}{2}\partial_{\mu}\partial_{\nu}\epsilon^{1}-\frac
{1}{2}\eta_{\mu\nu}\epsilon^{1},\text{ \ }(\partial^{2}-2)\epsilon
^{1}=0.\tag{4.55}%
\end{equation}
Eqs.(4.54) and (4.55) are consistent with the massive $\sigma$-model
calculation in the OCFQ string theory in \cite{7}.
\end{enumerate}

\qquad For the mass level $m^{2}=4$, by using Eqs.(4.24) and (4.25), the
linearized gauge transformation of WSFT in Eq.(4.5) gives
\begin{align}
\delta C_{\mu\nu\lambda}  &  =-\partial_{(\mu}\epsilon_{\nu\lambda)}^{0}%
-\frac{1}{2}\epsilon_{(\mu}^{2}\eta_{\mu\nu)},\tag{4.56}\\
\delta C_{[\mu\nu]}  &  =-\partial_{\lbrack\nu}\epsilon_{\mu]}^{1}%
-\partial_{\lbrack\mu}\epsilon_{\nu]}^{2},\tag{4.57}\\
\delta C_{(\mu\nu)}  &  =-\partial_{(\nu}\epsilon_{\mu)}^{1}-\partial_{(\mu
}\epsilon_{\nu)}^{2}+2\epsilon_{\mu\nu}^{0}-\epsilon^{2}\eta_{\mu\nu
},\tag{4.58}\\
\delta C_{\mu}  &  =-\partial_{\mu}\epsilon^{2}+2\epsilon_{\mu}^{1}%
+\epsilon_{\mu}^{2},\tag{4.59}\\
\delta\gamma_{\mu\nu}  &  =\frac{1}{2}(\partial^{2}-4)\epsilon_{\mu\nu}%
^{0}-\frac{1}{2}\epsilon^{3}\eta_{\mu\nu},\tag{4.60}\\
\delta\gamma_{\mu}^{0}  &  =\frac{1}{2}(\partial^{2}-4)\epsilon_{\mu}%
^{2}+\partial_{\mu}\epsilon^{3},\tag{4.61}\\
\delta\gamma_{\mu}^{1}  &  =-2\partial^{\nu}\epsilon_{\nu\mu}^{0}%
-2\epsilon_{\mu}^{1}-3\epsilon_{\mu}^{2},\tag{4.62}\\
\delta\gamma_{\mu}^{2}  &  =\frac{1}{2}(\partial^{2}-4)\epsilon_{\mu}%
^{1}-\partial_{\mu}\epsilon^{3},\tag{4.63}\\
\delta\gamma^{0}  &  =\frac{1}{2}(\partial^{2}-4)\epsilon^{2}-\epsilon
^{3},\tag{4.64}\\
\delta\gamma^{1}  &  =-\partial^{\mu}\epsilon_{\mu}^{2}-4\epsilon
^{2}-2\epsilon^{3},\tag{4.65}\\
\delta\gamma^{2}  &  =-2\partial^{\mu}\epsilon_{\mu}^{1}-5\epsilon
^{2}+4\epsilon^{3}+\epsilon_{\mu}^{0\mu}. \tag{4.66}%
\end{align}
For the gauge transformation induced by $D_{2}$ zero-norm state in Eq.(4.48),
for example, one can use Eqs.(4.39)-(4.46) with $C=-i/16$ to eliminate
Eqs.(4.60)-(4.66). One can then use the fact that background fields
$C_{(\mu\nu)}$ and $C_{\mu}$ are gauge artifacts of $C_{\mu\nu\lambda}$ in the
$\sigma$-model calculation, and deduce from Eqs.(4.56)-(4.59) the
inter-particle symmetry transformation
\begin{equation}
\delta C_{\mu\nu\lambda}=\frac{1}{2}\partial_{(\mu}\partial_{\nu}%
\epsilon_{\lambda)}^{(D_{2})}-2\eta_{(\mu\nu}\epsilon_{\lambda)}^{(D_{2}%
)},\text{ \ \ }\delta C_{[\mu\nu]}=9\partial_{\lbrack\mu}\epsilon_{\nu
]}^{(D_{2})}, \tag{4.67}%
\end{equation}
where $\partial^{\lambda}\epsilon_{\lambda}^{(D_{2})}=0,(\partial
^{2}-4)\epsilon_{\lambda}^{(D_{2})}=0$. The other three gauge transformations
corresponding to three other zero-norm states, the spin-two, $D_{1}$, and
scalar can be similarly constructed from Eqs.(4.56)-(4.66). One gets%
\begin{equation}
\delta C_{\mu\nu\lambda}=\partial_{(\mu}\epsilon_{\nu\lambda)};\text{
\ }\partial^{\mu}\epsilon_{\mu\nu}=0,\text{ }(\partial^{2}-4)\epsilon_{\mu\nu
}=0, \tag{4.68}%
\end{equation}%
\begin{equation}
\delta C_{\mu\nu\lambda}=\frac{5}{2}\partial_{(\mu}\partial_{\nu}%
\epsilon_{\lambda)}^{(D_{1})}-\eta_{(\mu\nu}\epsilon_{\lambda)}^{(D_{1}%
)};\text{ \ }\partial^{\lambda}\epsilon_{\lambda}^{(D_{1})}=0,\text{
}(\partial^{2}-4)\epsilon_{\lambda}^{(D_{1})}=0, \tag{4.69}%
\end{equation}%
\begin{equation}
\delta C_{\mu\nu\lambda}=\frac{17}{4}\partial_{\mu}\partial_{\nu}%
\partial_{\lambda}\theta-\frac{9}{2}\eta_{(\mu\nu}\theta_{\lambda)};\text{
}(\partial^{2}-4)\theta=0. \tag{4.70}%
\end{equation}
Eqs.(4.67)-(4.70) are exactly the same as those calculated by the generalized
massive $\sigma$-model approach of string theory \cite{7}.

We thus have shown in this section that off-shell gauge transformations of
WSFT are identical to the on-shell stringy gauge symmetries generated by two
types of zero-norm states in the OCFQ string theory. The high energy limit of
these stringy gauge symmetries generated by zero-norm states was recently used
to calculate the proportionality constants among high energy scattering
amplitudes of different string states conjectured by Gross \cite{5}.
\textit{Based on the zero-norm state calculations in \cite{1,2,3} and the
calculations in this section, we thus have related gauge symmetry of WSFT
\cite{15} to the high-energy stringy symmetry conjectured by Gross
\cite{4,5,6}.}

\section{Conclusion}

In this paper, we have calculated zero-norm states in the OCFQ string theory,
the light-cone DDF string theory and the off-shell BRST string theory. In the
OCFQ string theory, we have solved the Virasoro constraints for all physical
states ( including zero-norm states) in the helicity basis. Much attention is
paid to discuss the inter-particle zero-norm state at the mass level $m^{2}%
=4$. We found that\ one can use polarization of either one of the two
positive-norm states to represent the polarization of the inter-particle
zero-norm state. This justifies how one can have the inter-particle symmetry
transformation for the two massive modes in the weak field massive $\sigma
$-model calculation derived previously \cite{7}. This inter-particle symmetry
transformation, in contrast to the high energy symmetry of Gross \cite{5}, is
valid to all energy.

In the light-cone DDF string theory, one can easily write down the general
formula for all zero-norm states in the spectrum. We have identified type I
and Type II zero-norm states up to the mass level $m^{2}=4.$ The analysis can
be easily generalized to any higher mass level as well.

Finally, we have calculated off-shell zero-norm states in the WSFT. After
imposing the no ghost conditions, we can recover two types of on-shell
zero-norm states in the OCFQ string theory. We then show that off-shell gauge
transformations of WSFT are identical to the on-shell stringy gauge symmetries
generated by two typse of zero-norm states in the generalized massive $\sigma
$-model approach of string theory. \textit{Based on these zero-norm state
calculations, we have thus related gauge symmetry of WSFT \cite{15} to the
high-energy stringy symmetry of Gross \cite{5}.}

\end{document}